\documentclass[%
 reprint,amsmath,amssymb,
 aps]{revtex4-1}

\usepackage[utf8]{inputenc}
\usepackage[T1]{fontenc} 

\usepackage{graphicx}
\usepackage{float} 
\usepackage{enumerate}

\usepackage{bm}
\usepackage{amsmath,amsfonts}
\usepackage{amssymb}
\usepackage{commath}
\usepackage{lipsum}

\usepackage{caption}
\usepackage[colorinlistoftodos]{todonotes} 

\newcommand{\Rop}{\hat{R}}
\newcommand{\Gop}{\hat{G}}
\newcommand{\Nop}{\hat{N}}

\newcommand{\Bop}{\hat{B}}
\newcommand{\Dop}{\hat{D}}
\newcommand{\Uop}{\hat{U}}
\newcommand{\Sop}{\hat{S}}
\newcommand{\Rmat}{\mathcal{R}}
\newcommand{\Bmat}{\mathcal{B}}

\newcommand{\Smat}{\mathcal{S}}
\newcommand{\rop}{\hat{\textbf{$\rho$}}}
\newcommand{\Jz}{\hat{J}_z}
\newcommand{\Jx}{\hat{J}_x}
\newcommand{\Jy}{\hat{J}_y}

\newcommand{\aop}{\hat{a}_1}
\newcommand{\adag}{\hat{a}_1^{\dagger}}
\newcommand{\bop}{\hat{a}_2}
\newcommand{\bdag}{\hat{a}_2^{\dagger}}
\newcommand{\iop}{\hat{a}_i}
\newcommand{\idag}{\hat{a}_i^{\dagger}}

\newcommand{\uvect}{\textbf{u}}
\newcommand{\xivect}{\bm{\xi}}
\newcommand{\ddagvect}{\bm{d}^{\dagger}}
\newcommand{\dvect}{\bm{d}}

\newcommand{\ket}[1]{|#1\rangle}
\newcommand{\bra}[1]{\langle #1|}
\newcommand{\moy}[1]{\langle #1 \rangle}

\begin{document}

\preprint{APS/123-QED}

\title{Metrological advantage at finite temperature for Gaussian phase estimation}

\author{Louis Garbe}
\author{Simone Felicetti}
\author{Perola Milman}
\author{Thomas Coudreau}
\author{Arne Keller}
\affiliation{Laboratoire Matériaux et Phénomènes Quantiques, Université Paris Diderot, CNRS UMR 7162, Sorbonne Paris Cité, France}

\date{\today}
 
\begin{abstract} 
In the context of phase estimation with Gaussian states, we introduce a quantifiable definition of metrological advantage that takes into account thermal noise in the preparation procedure. For a broad set of states, \textit{isotropic non-pure Gaussian states}, we show that squeezing is not only necessary, but sufficient, to achieve metrological advantage. We interpret our results in the framework of resource theory, and discuss possible sources of advantage other than squeezing. Our work is a step towards using phase estimation with pure and mixed state to define and quantify nonclassicality. This work is complementary with studies that defines nonclassicality using quadrature displacement estimation.
\end{abstract}

\maketitle

\section{Introduction}

In recent years, advances in our understanding and capacity to manipulate quantum states have opened new promising technological avenues, especially in the field of computation, communication, or metrology. In metrology especially, it was already suggested in the 80s \cite{caves_quantum-mechanical_1981} that non-classical correlations between photonic probes could be used to reduce the photon counting error, or shot-noise. Such \textit{metrological advantages} are now firmly established, both theoretically and experimentally, on various platforms \cite{hosten_measurement_2016,ligo_scientific_collaboration_enhanced_2013,demkowicz-dobrzanski_chapter_2015,pezze_quantum_2014,pezze_quantum_2018}. For instance, a phase estimation protocol involving $N$ uncorrelated probes will be limited by an error scaling at most like $\frac{1}{\sqrt{N}}$, the so-called \textit{Standard Quantum Limit} (SQL).  With quantum-correlated probes, however, it is possible to beat this limit and to reach the \textit{Heisenberg limit} $\frac{1}{N}$. If the number of probes is fluctuating (which is the case, \textit{e.g.}, if light is used in an interferometry experiment), the SQL bound is $\frac{1}{\sqrt{\moy{\Nop}}}$, with $\Nop$ the total particle number operator.

Reversing the perspective, the ability to perform tasks intractable by classical means can also be used to define nonclassicality. In particular, several recent works have studied how metrological advantage could be used as a quantifier of nonclassicality of continuous-variable states \cite{kwon_nonclassicality_2018,yadin_operational_2018}. These nonclassicality quantifier can be used, for instance, to establish formal limitations on state preparation procedures. These works have focused on quadrature displacement estimation, which involve \textit{linear} combination of the field quadratures.
Experimentally, however, the most relevant metrological protocols are phase estimation protocols such as interferometric schemes, which involve \textit{quadratic} combination of  the field quadratures. This begs the questions: what is the link between nonclassicality and phase estimation advantage ? Could the latter be used to quantify the former?

To gain intuition on these challenging issues, it is essential to understand precisely which states allow to achieve metrological advantage. An important step in this direction was made in \cite{hyllus_not_2010}, where, for fixed number of probes, the set of pure states which achieve phase estimation advantage has been completely characterized. Significantly, it was found that entanglement was not sufficient to guarantee an advantage. To the best of our knowledge, however, no such characterization has been made in the case of a fluctuating number of probes and noisy protocols, which is relevant for most experiments.

The presence of noise in the protocol is a particularly significant issue. Numerous works \cite{escher_general_2011,alipour_quantum_2014,spagnolo_phase_2012,demkowicz-dobrzanski_adaptive_2017,zhou_achieving_2018,albarelli_restoring_2018,nichols_multiparameter_2017}  have studied the maximum precision attainable in noisy  interferometric scheme (for instance, when photon loss is present in the interferometer). It was proven that Heisenberg limit becomes unattainable under very general noise condition. Similarly, it is critical to consider imperfect state preparation by studying mixed input states in addition to pure ones. In this case, however, metrological advantage become harder to define, since even classical states can not reach the SQL any more.

In this work, we tackled this issue by explicitly introducing preparation imperfection in the definition of metrological advantage. We studied this quantity for a broad class of mixed photonic input states, \textit{isotropic Gaussian states}. We found that squeezing is both necessary and sufficient to achieve metrological advantage. This manuscript is organized as follows: in Section \ref{secsetup}, we introduce the protocols and states we considered. In Section \ref{Secftql}, we introduce the notion of \textit{Finite Temperature Quantum Limit} or FTQL, which is a generalization of SQL taking thermal noise in the preparation stage into account. We characterize the states that achieve sub-FTQL error, showing that squeezing is both necessary and sufficient to beat this limit. We also show this property is no longer true in single-mode interferometry. In Section \ref{secdiscussion}, we discuss the interpretation of our findings in two different contexts. First, we study whether our work may be rephrased in the \textit{resource theory} framework \cite{brandao_second_2015,baumgratz_quantifying_2014}. We find many of our results fit in this framework ; however, there are also interesting differences. Second, we review other possible sources of metrological advantage, \textit{particle entanglement} \cite{demkowicz-dobrzanski_chapter_2015,braun_quantum-enhanced_2018} and photon number superposition, and their link with squeezing and our results. Finally, we conclude in Section \ref{secconclusion}.


\section{Setup}
\label{secsetup}

\subsection{Protocol}

\begin{figure}
		\includegraphics[scale=0.2]{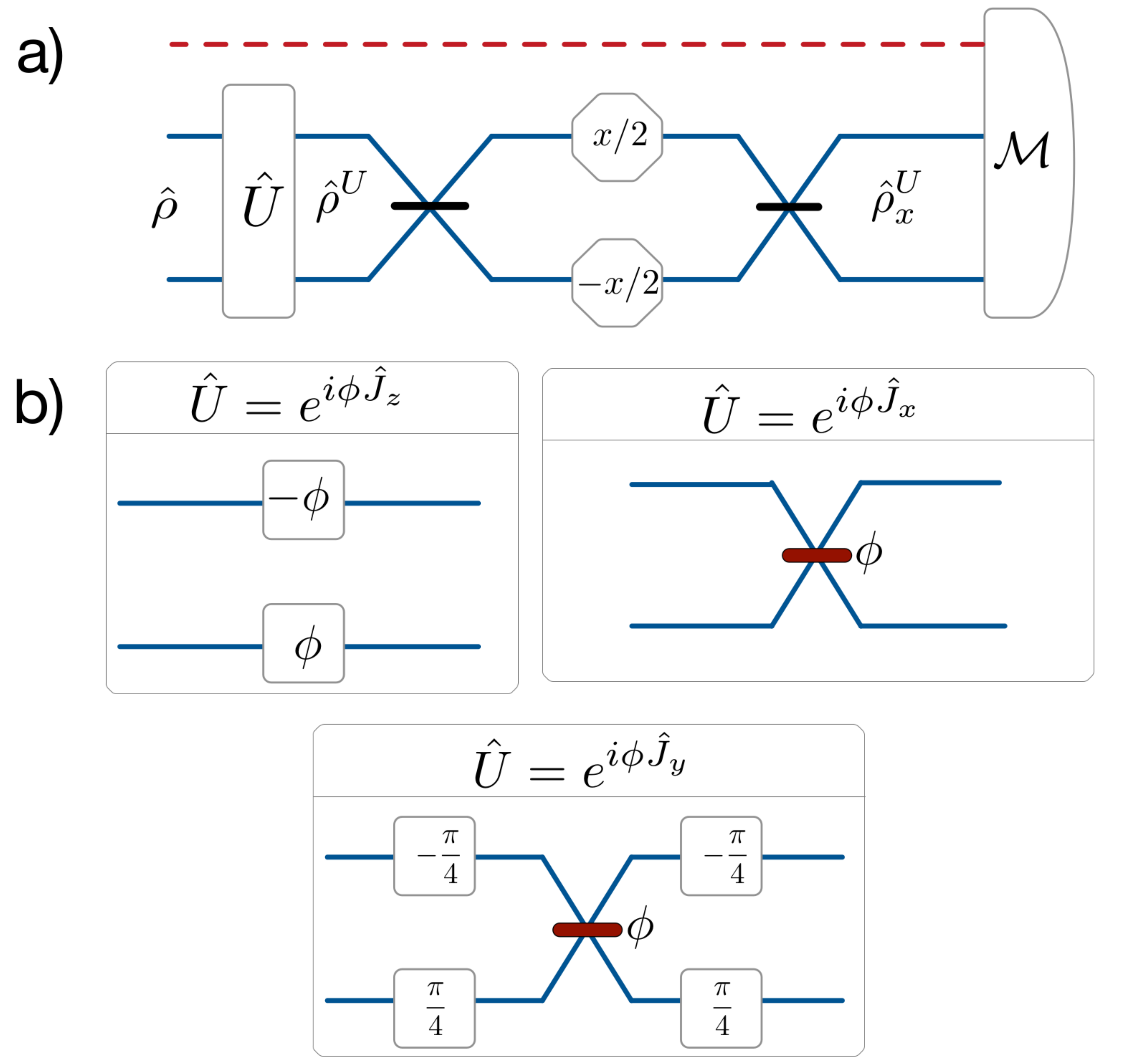}
	\caption{a) Schematic depiction of the metrological setup. The input state $\rop$ is modified by a unitary $\Uop$, sent inside the interferometer, then measured. We also consider a perfect phase reference (red dotted line) which allows to make homodyne measurements. b) Different possible configurations for $\Uop$. $\Jz$ rotations are only phase differences. A rotation $e^{i\phi\Jx}$ corresponds to a beam splitter with transmission coefficient $\cos{\frac{\phi}{2}}$ (which can also be emulated with phase-shifters and balanced beam-splitters). The Mach-Zehnder in a) is described by a rotation $e^{i\phi\Jy}$; such an operation can also be directly implemented with an unbalanced beam-splitter as displayed in b) . In all figures, we have omitted additional $\frac{\pi}{2}$ phase shifts for readability.}
	\label{figinterfero}
\end{figure}


We consider the estimation of a phase difference $x$ between both arms of an optical Mach-Zehnder interferometer.
This simple but paradigmatic setup is relevant for various experimental protocols such as gravitational wave detection \cite{ligo_scientific_collaboration_enhanced_2013,adhikari_gravitational_2014}, and the same formalism may be used in atoms-based metrology \cite{cronin_optics_2009,pezze_quantum_2018,hosten_measurement_2016}. It has also been argued \cite{knott_local_2016} that some multi-channel protocols can be reduced to a series of two-channel interferometers. We will allow for all measurements, including homodyne measurements on both arms, which require a perfect phase reference.\\

 Let us note first that with different configurations of the interferometer (for instance, different reflexivity of the beam-splitters), the precision of the estimation may be different. We model these changes by the addition of a unitary $\Uop$ before the interferometer: See \figref{figinterfero}. We will consider $\Uop$ that are Passive Linear Operation (PLO), \textit{i.e.}, that can be implemented using only phase-shifters and beam-splitters.

 Note that one-mode Gaussian phase estimation can also be studied in this setting; it corresponds to sending the vacuum in one arm, removing the beam-splitter (which can also be described by adding a suitable $\Uop$), and performing homodyne detection.

 PLOs are readily available experimentally; thus, any change of $\Uop$ resulting in an increase in precision could hardly be considered a metrological advantage. To single out the effects of the input state properties on the estimation precision, we will optimize the precision over $\Uop$. \\

Formally, one can define the operators $\Jz=\frac{\adag\aop-\bdag\bop}{2}$, $\Jx=\frac{\adag\bop+\aop\bdag}{2}$, and $\Jy=i\frac{\adag\bop-\aop\bdag}{2}$, where $\iop$ ($\idag$) is the photon annihilation (creation) operator in the $i$-th arm (with $i=1,2$). These quantities verify the commutation relation $[\hat{J}_a,\hat{J}_b]=i\epsilon _{abc}\hat{J}_c$, with $\epsilon _{abc}$ the totally antisymmetric tensor. PLOs can be described as rotations generated by these operators: see \figref{figinterfero}. Without loss of generality, we may write $\Uop=e^{i\textbf{u}.\hat{\textbf{J}}}$ with $\hat{\textbf{J}}=(\Jx,\Jy,\Jz)$ and $\textbf{u}=(u_x,u_y,u_z)$. The Mach-Zehnder interferometer may be described by a rotation $e^{-ix \Jy}$ (for the sake of clarity, we have omitted here additional $\frac{\pi}{2}$ phase-shifts which have no influence on the analysis). \\

Thus, a state $\rop_0$ sent into the interferometer evolves into the state: 
\begin{equation}
	\rop_{x }^{\Uop}=e^{-ix \Jy}\Uop\rop_0\Uop^{\dagger}e^{ix \Jy}
\end{equation}
Measuring observables on $\rop_{x }^{\Uop}$ then allows to estimate the phase $x$. When optimizing the choice of observable and parameter reconstruction, and repeating the experiment independently $n$ times, the estimation error, $\delta_x$, is bounded by the Quantum Cramer-Rao bound: 
\begin{equation}
	\delta x \geq \frac{1}{\sqrt{n}\sqrt{I_F(\rop_{x }^{\Uop})}}
\end{equation}
 where $I_F(\rop_{x }^{\Uop})$ is the Quantum Fisher Information (QFI), defined as:
\begin{equation}
	I_F(\rop_{x }^{\Uop})=Tr\Big((\hat{L}_{x }^{\Uop})^2\rop_{x }^{\Uop}\Big)
\end{equation}

$L_{x }^{\Uop}$ is the symmetric logarithmic derivative, defined as:
\begin{equation}
\label{defL}
	\dot{\rop}_{x }^{\Uop}=\frac{\hat{L}_{x }^{\Uop}\rop_{x }^{\Uop}+\rop_{x }^{\Uop} \hat{L}_{x }^{\Uop}}{2}
\end{equation}
where the dot denotes derivation with respect to $x $. Note again that the QFI is meaningful only if all measurements are allowed, including homodyne measurements.

For one-parameter estimation, the Quantum Cramer-Rao bound is attainable. We then define the \textit{metrological capacity} of a state $\rop_0$ as the best precision one can achieve by optimizing the PLO operation $\Uop$. Formally:

\begin{equation}
	I_F^{\text{opt}}(\rop_0)=\text{Max}_{\{\Uop\}} I_F(\rop_{x }^{\Uop})
\end{equation}
Note that this quantity may also be seen as the maximum eigenvalue of the QFI matrix \cite{genoni_optimal_2013,ragy_compatibility_2016,nichols_multiparameter_2017} corresponding to the simultaneous estimation of three rotation angles around the $\Jz$, $\Jx$, $\Jy$ axis. \footnote{Note also that optimization over $\Uop$ may be seen either as optimization of the input state, or optimization of the operator encoding phase $e^{-i x \Uop^{\dagger}\Jx\Uop}$}. \\

Since (\ref{defL}) is an implicit equation, finding $\hat{L}_{x }^{\Uop}$ and $I_F$ is generally a highly non-trivial task; in all generality, finding the QFI is as difficult as state diagonalization. The problem, however, can become tractable for specific classes of states. The best-known of such states are Gaussian states, \textit{i.e.}, states with Gaussian Wigner function. Because of their experimental availability, a vast body of work has been published on Gaussian states in the context of metrology: see, \textit{i.e.}, \cite{pinel_quantum_2013,schmeissner_spectral_2014,monras_phase_2013,jiang_quantum_2014,safranek_gaussian_2016, safranek_optimal_2016,safranek_quantum_2015-1, sparaciari_bounds_2015,sparaciari_gaussian-state_2016}. In what follows, we will consider Gaussian input $\rop$.

\subsection{Parameterization of Gaussian states} 
For our purpose, it is paramount to find a proper parameterization of the input $\rop$; this is the aim of the present subsection. 

Any Gaussian state may be obtained by applying so-called Gaussian operations on a thermal state \cite{simon_quantum-noise_1994,safranek_gaussian_2016}:

\begin{equation}
\label{symplecform}
	\rop_0=\Gop\rop_{\text{th}}^{(1)}(T_1)\otimes\rop_{\text{th}}^{(2)}(T_2)\Gop^{\dagger}
\end{equation}
where $\Gop$ is a Gaussian (unitary) operation, and $\rop_{\text{th}}^{(i)}(T_i)=\frac{1}{Z_i}e^{-\frac{\hbar\omega_i \idag\iop}{kT_i}}$, with $k$ the Boltzmann constant, $T_i$ the temperature, $Z_i$ the partition function and $\omega_i$ the frequency of the $i$-th mode. Gaussian operations include displacement, squeezing, mode-mixing, phase-shifting, and composition thereof. Any Gaussian operation can be decomposed as: 
\begin{equation}
\label{symplecdef}
	\begin{split}
	\Gop=\Dop(\bm{\gamma})\Rop_1(\phi_1)\Rop_2(\phi_2)\Bop(\theta)\Rop_{as}(\psi) \\
	 \Sop_1(r_1)\Sop_2(r_2)\Rop_{as}(\psi_1)\Bop(\theta_1)
	\end{split}
\end{equation}

Here $\Dop$ is a displacement operation $\Dop(\bm{\gamma})=e^{\bm{\gamma}.\hat{\textbf{a}}^{\dagger}-\overline{\bm{\gamma}}.\hat{\textbf{a}}}$, where $$\hat{\textbf{a}}=(\aop,\bop)^T$$ and the bar denotes complex conjugation. $\Rop_i(\phi)=e^{i\phi\idag\iop}$ denotes phase-shifting of the $i$-th mode, $\Bop(\theta)=e^{2i\theta\Jy}$ stands for mode-mixing, $\Rop_{as}(\psi)=\Rop_1(\psi)\Rop_2(-\psi)=e^{i2\psi\Jz}$, and $\Sop_i(r_i)=e^{\frac{r_i}{2}\Big((\iop)^2-(\idag)^2\Big)}$ represents squeezing of mode $i$ by a real parameter $r_i$. We also define $\alpha$, $\phi_{d1}$, $\phi_{d2}$ such that: $$\bm{\gamma}=\lvert\gamma\rvert ( e^{i\phi_{d1}} \cos{\alpha}, e^{i\phi_{d2}} \sin{\alpha})$$.\\

The thermal states $\rop_{\text{th}}^{(i)}(T_i)$ are fully parameterized by the parameters: 

$$\nu_i=\text{cotanh}\Big(\frac{\hbar\omega_i}{2kT_i}\Big)=\frac{1}{\text{Tr}\Big((\rop_{\text{th}}^{(i)})^2\Big)}$$  

We refer to these as \textit{symplectic eigenvalues} (See Appendix \ref{appendixA} for more details). We have $\nu_i\geq1$, with equality if and only if $T_i=0$. The set of symplectic eigenvalues associated with a given Gaussian state is unique.

Note that we have $\text{Tr}\Big(\rop_0^2\Big)=\frac{1}{\nu_1\nu_2}$. Hence, the  impurity of a two-mode Gaussian state can be seen as thermal noises on both modes (this also holds true for multi-modes Gaussian states).\\

We will impose the further condition $\omega_1=\omega_2$ and $T_1=T_2=T$ (and hence $\nu_1=\nu_2=\nu$). Such states we call \textit{isotropic} Gaussian states. Those states are relevant when the two arms are submitted to identical and independent noise sources, which is well motivated experimentally. For these states, we may set $\theta_1=\psi_1=0$ without loss of generality (See Appendix \ref{appendixA}). \\

Hence, all isotropic Gaussian states can be described by the following set of parameters:
 $(\nu,\lvert\gamma\rvert,\alpha,\phi_{d1},\phi_{d2},\phi_1,\phi_2,\theta,\psi,r_1,r_2)$. States with $r_1=r_2=0$ are \textit{displaced thermal states}; we can generically call all the other states \textit{squeezed states}. Note that this definition of squeezing is not exactly identical to the usual quantity defined using quadrature variance (or spin variance for atoms, as measured by the Wineland criterion \cite{wineland_spin_1992}). Rather, squeezed states are defined here as states that cannot be generated using only thermal states, displacement operations and PLOs. This definition of squeezing was first put forward by Simon, Mukunda, and Dutta in \cite{simon_quantum-noise_1994}. \\

Applying the operator $\Uop$ then amounts to changing the parameters of the input state with some constraints, the most prominent of which being that the total photon number $\moy{\Nop}$ is conserved. For completeness, we give here the expression of this number: $\langle\Nop\rangle = \langle\adag\aop+\bdag\bop\rangle = \nu(\sinh(r_1)^2+\sinh(r_2)^2)+\lvert \mathbf{\gamma}\rvert^2 + (\nu-1)$. For a displaced thermal state, this reduces to $\lvert \mathbf{\gamma}\rvert^2 + (\nu-1)$.

\section{FTQL and metrological advantage} 
\label{Secftql}

\subsection{Definition of FTQL and metrological advantage}
To have a working notion of quantum metrological advantage, it is key to properly identify an ensemble of "classical" states. The maximum precision attainable using those states serves as a threshold upon which the notion of advantage may be built. For (isotropic) Gaussian states, displaced thermal states are good candidates to define the reference.

Usually, the metrological advantage would be defined by taking a pure coherent state with the same average number of photons $\moy{\Nop}$ as the state $\rop_0$ under investigation. Such a state has a metrological capacity $I_F^{\text{opt}}=4\moy{\Nop}=4\text{Tr}(\Nop\rop_0)$: this value is the SQL. Then, we can say that a state $\rop_0$ achieve metrological advantage if $I_F(\rop_0)\geq 4\text{Tr}(\Nop\rop_0) $, irrespective of its temperature. This definition is relevant, for instance, when the average number of photons is fixed by experimental constraints, but thermal noise is not. Then, we may act on both squeezing and thermal noise to achieve higher-than-SQL precision.

By contrast, here, we are going to include explicitly the thermal noise in the definition of the reference. Specifically, for every state $\rop_0$, the reference state is an isotropic displaced thermal state $\hat{h}_{\text{ref}}(\rop_0)$ with the same $\moy{\Nop}$ \textbf{and} the same $\nu$ as $\rop_0$. This definition is particularly relevant in situations where both average photon number and thermal noise are set by external constraints.
The reference state has a metrological capacity $I_F^{\text{opt}}(\hat{h}_{\text{ref}})=\frac{4\lvert\gamma\rvert^2}{\nu}$ \cite{safranek_gaussian_2016}. Reexpressing this as a function of the number of photons in the reference state, we find $I_F^{\text{opt}}(\hat{h}_{\text{ref}})=4\frac{\langle \Nop\rangle+1}{\nu}-4$. This bound we call the \textit{Finite-Temperature Quantum Limit} or FTQL. This precision increases with the number of photons, and decreases with the temperature \footnote{Note that $\hat{h}_{\text{ref}}(\rop_0)$ is not uniquely defined, since there are infinitely many displaced thermal states with the same $<\hat{N}>$ and $\nu$. However, for the kind of interferometers we consider, displaced thermal states with identical $\nu$ and $<\Nop>$ can be mapped to each other by PLOs, and hence they all achieve the same metrological capacity (actually, for a displaced thermal state $\hat{h}$, $I_F(\hat{h}^U)=I_F(\hat{h})$ for all PLOs $\Uop$, which also means the protocol needs not be optimized for those states). Thus, the FTQL is well-defined.}. Note that FTQL and SQL become equivalent in the limit of zero temperature (\textit{i.e.} $\nu=1$). In what follows, we will use the shorter notation $I_F^{\text{ref}}(\rop_0)=I_F^{\text{opt}}(\hat{h}_{\text{ref}}(\rop_0))$.

To summarize, for an arbitrary isotropic two-mode Gaussian state $\rop_0$ with symplectic eigenvalue $\nu$, we define metrological advantage as:

\begin{eqnarray}
\label{metroadvFTQL}
	\mathcal{A}_G(\rop_0) & = \text{Max}\Big( I_F^{\text{opt}}(\rop_0)-I_F^{\text{ref}}(\rop_0) , 0 \Big) \\ \nonumber
	 & = \text{Max}\Big( I_F^{\text{opt}}(\rop_0)-4\frac{\langle \Nop\rangle+1}{\nu}+4 , 0 \Big)
\end{eqnarray}

\subsection{Main results}

We are now going to study under which conditions we have $\mathcal{A}_G$$ > 0$. We have proven the following theorem:\\ 

\textbf{Theorem 1} \textit{Let $\rop_0$ be an isotropic non-pure two-mode Gaussian state. Then $\mathcal{A}_G(\rop_0)$$=0$ if and only if $\rop_0$ is a displaced thermal state.}\\

The sufficient part of this statement is trivial, since all displaced thermal states have $\mathcal{A}_G(\rop_0)=0$ by definition. This means that given the way we have defined metrological advantage, squeezing is necessary to achieve it. What is nontrivial is the converse: if a non-pure state is squeezed, then it has nonzero metrological advantage. If squeezing is present, no matter the amount or direction, the beam-splitters and phase-shifters of $\Uop$ always give us enough control to exploit it. This theorem, along with our definition of metrological advantage, constitutes our main result.\\

The complete proof of this theorem is provided in the Appendix, we only give a sketch here. 

We consider a state $\rop_0$ described by $(\nu,\lvert\gamma\rvert,\alpha,\phi_{d1},\phi_{d2},\phi_1,\phi_2,\theta,\psi,r_1,r_2)$. We will set $\theta=0$ and $\psi=0$, which means we do not have two-mode squeezing: we prove in the Appendix \ref{appendixC} that all states are amenable to this form using PLOs. We made extensive use of \cite{safranek_gaussian_2016}, which provides a complete expression for the QFI of a Gaussian state, and allowed us to rewrite the advantage as:

\begin{eqnarray}
	\label{maineq}
	\nonumber
	I_F(\rop_{x }^{\Uop})-I_F^{\text{ref}}(\rop_0) & = 2[o^2 X +p^2Y] + \frac{4\lvert\gamma\rvert^2}{\nu}\Big(e^{2r_1}\kappa^2 \\ \nonumber
	  & + e^{-2r_1}\delta^2 + e^{2r_2}\upsilon^2 + e^{-2r_2}\lambda^2 -1 \Big)\\
	= & 2[o^2 X +p^2Y] + \frac{4\lvert\gamma\rvert^2}{\nu}V
\end{eqnarray}

The general expressions of all the parameters are given in Appendix \ref{appendixB}. It will suffice here to give their values when $\theta=\psi=0$: $o=\sin(\phi_1-\phi_2)$, $p=-\cos(\phi_1-\phi_2)$, $\kappa=\sin(\alpha)\cos(\phi_1+\phi_{d2})$, $\delta=\sin(\alpha)\sin(\phi_1+\phi_{d2})$, $\upsilon=\cos(\alpha)\cos(\phi_2+\phi_{d1})$, $\lambda=\cos(\alpha)\sin(\phi_2+\phi_{d1})$. $X$ and $Y$ depend only on $\nu$ and the squeezing parameters, and verify $X \geq \lvert Y \rvert \geq 0$. Thus, there are two contributions to the metrological advantage: the first one comes from squeezing only, the second one from the interplay between squeezing and displacement. Now, our goal is to understand how, using a PLO operator $\Uop$ on the state $\rop_0$, we may modify the values of the various parameters so as to render Eq.\eqref{maineq} positive.

We will first assume $\lvert\gamma\rvert=0$. A simple phase-shifting operation allows to set $\phi_1-\phi_2=\frac{\pi}{2}$, which gives $p=0$ and $o=1$; this immediately yields $I_F-I_F^{\text{ref}}=2X\geq0$. The inequality is strict as soon as $r_1\neq0$ or $r_2\neq0$. Thus, we achieve advantage simply by applying a phase shift to our state prior to sending it in the interferometer. Now, if $\lvert\gamma\rvert\neq0$, two cases are possible, depending on the value of $V$ in \eqref{maineq}:
 
 \begin{itemize}
\item If $V > 0$, $I_F-I_F^{\text{ref}}$ is positive for all value of $\lvert\gamma\rvert$. 
\item If $V\leq 0$, $I_F-I_F^{\text{ref}}$ becomes negative as soon as $\lvert\gamma\rvert$ crosses a certain threshold $\lvert\gamma_s\rvert$. In that case, we adapt the phase-shifting to the value of the displacement; if $\lvert\gamma\rvert\leq\lvert\gamma_s\rvert$, we keep the same parameter values. If $\lvert\gamma\rvert\geq\lvert\gamma_s\rvert$, we apply an additional phase shift $\Uop=\Rop_{as}(\chi)$. This sets new values for the parameters: $\phi_1\rightarrow\phi_1 + \chi$, $\phi_2\rightarrow\phi_2 - \chi$, $\phi_{d1}\rightarrow\phi_{d1} - \chi$ (pay attention to the signs in the definition of $\Rop$ and $\phi_{d1}$) and $\phi_{d2}\rightarrow\phi_{d2} + \chi$. The other parameters remain constant. A short calculation shows that, for $\chi=\frac{\pi}{4}$, $V$ becomes positive; using $\lvert\gamma\rvert\geq\lvert\gamma_s\rvert$, we can readily show that the displacement-squeezing terms dominates the pure squeezing term in absolute value, which entails that $I_F-I_F^{\text{ref}}$ is positive. 
\end{itemize}

To summarize, we have a pure squeezing term and a displacement-squeezing term. Depending on the value of $\lvert\gamma\rvert$, one term will dominate the other (in absolute value). We have shown how, using PLOs, we can ensure that the dominant term will always be positive.\\

This result is valid for non-pure states. However, it also holds true for most pure states. The sole exceptions to that rule are some very specific pure squeezed states for which the FTQL may be reached, but not surpassed; see Appendix \ref{appendixD} and \ref{appendixE} for further details.\\ 

\subsection{Exemples}

Consider states of the form $$\rop_i= \Dop(\bm{\gamma})\Rop_1(\phi_1)\Rop_2(\phi_2)\Sop_1(r)\rop_{\text{th}}(\nu)(.)^{\dagger}$$

This means the first arm of the interferometer holds a displaced squeezed state, while the second holds a thermal state. The relative orientation of squeezing and displacement are given by the parameter $\phi_1+\phi_d$.

For those states, we found explicitly the optimal protocol and $I_F^{\text{opt}}$ (see Appendix \ref{appendixD}). In \figref{courbes}, we have plotted the evolution of $\frac{I_F^{\text{opt}}-I_F^{\text{ref}}}{I_F^{\text{ref}}}$ as a function of displacement $\lvert\gamma\rvert$, squeezing $r$, and temperature $\nu$. For the sake of clarity, we display only two cases: $\phi_1+\phi_d=0$ (displacement and noise reduction in the same direction) and $\phi_1+\phi_d=\frac{\pi}{2}$ (displacement and noise reduction in orthogonal directions).

In the case $\phi_1+\phi_{d}=0$, the curves show two distinct regimes; these regimes are characterized by $-4\frac{\lvert\gamma\rvert^2 (e^{-2r}-1))}{\nu} - \frac{4\nu^2}{\nu^2+1}\big(\sinh^2(2r)-2\sinh^2(r_1)\big)$ $\leq0$ or $\geq0$. Each of these regimes corresponds to a different optimal strategy (Appendix \ref{appendixD}). By contrast, in the case $\phi_1+\phi_{d}=\frac{\pi}{2}$, the optimal strategy is unique.

First, we see that $\frac{I_F^{\text{opt}}-I_F^{\text{ref}}}{I_F^{\text{ref}}}$ tends to decrease with $\lvert\gamma\rvert$, before stabilizing. In the case $\phi_1+\phi_{d}=0$, the advantage vanishes when $\lvert\gamma\rvert$ goes to infinity. For $\phi_1+\phi_{d}=\frac{\pi}{2}$ however, we retain an advantage for even for very large $\lvert\gamma\rvert$. Second, the advantage increases with squeezing in both cases, which was to be expected. Finally, $\frac{I_F^{\text{opt}}-I_F^{\text{ref}}}{I_F^{\text{ref}}}$ also increases with the temperature. This is less intuitive, since thermal noise is usually considered detrimental for metrological estimation. It is known, however, that squeezed thermal state can actually measure a phase more accurately than pure squeezed states \cite{safranek_optimal_2016}.

Finally, to better understand the significance of our result, we made a comparison with the one-mode case. In Appendix \ref{appendixE} , we found the explicit expression of $I_F^{\text{opt}}$ for an arbitrary one-mode state (as for the two-mode estimation, we allow both photon-counter and homodyne measurement). Very interestingly, some one-mode squeezed states perform worse than displaced thermal state for phase estimation. This difference between one- and two-mode estimation comes from the additional degree of control available with two modes: one-mode states may only be transformed by phase-shifters, while two-mode states can also be acted upon by beam-splitters. This additional degree of control is the reason why squeezing can always be put to good use in the two-mode case, but not in the one-mode case.

\begin{figure}
	\hspace{-18pt}\begin{tabular}{| p{.25\textwidth} | p{.25\textwidth}|}
	\hline
	 \begin{center}\includegraphics[width=.2\textwidth]{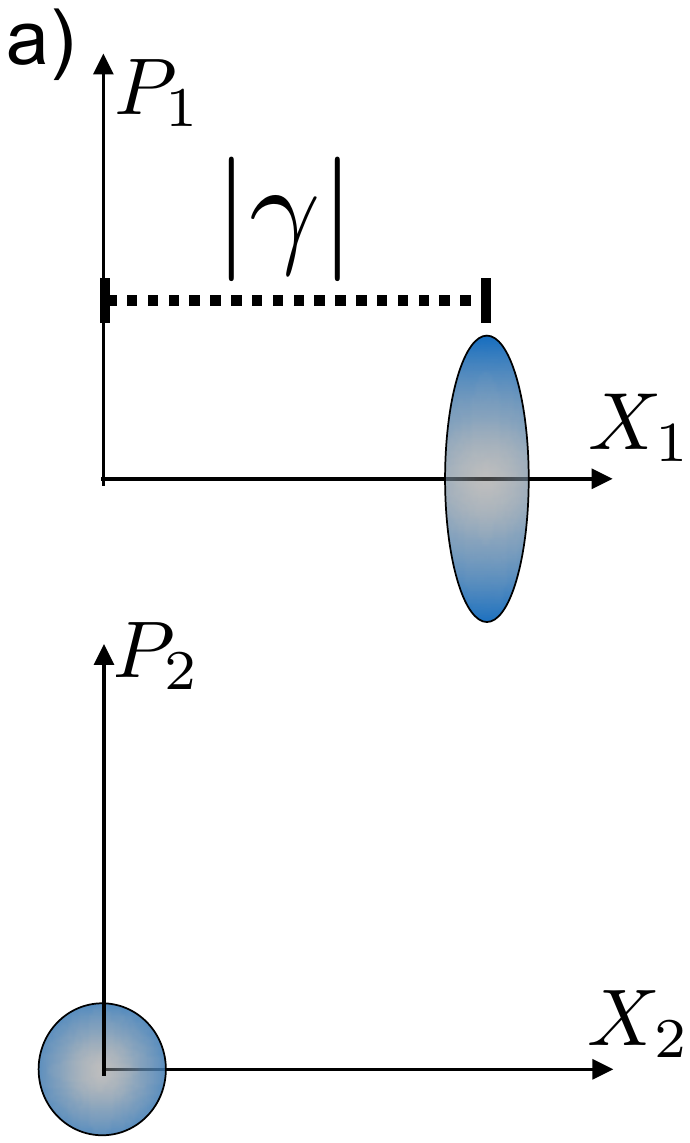}\end{center}	&  \begin{center}\includegraphics[width=.2\textwidth]{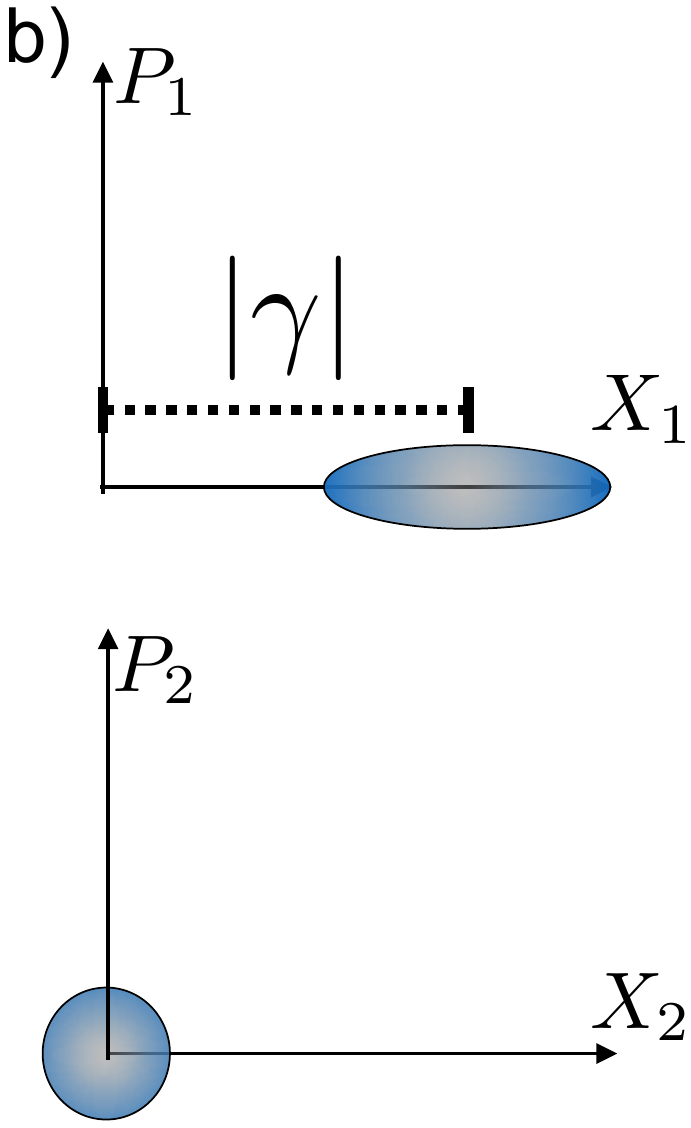}\end{center} \\
	\hline
	 \begin{center}\includegraphics[width=.25\textwidth]{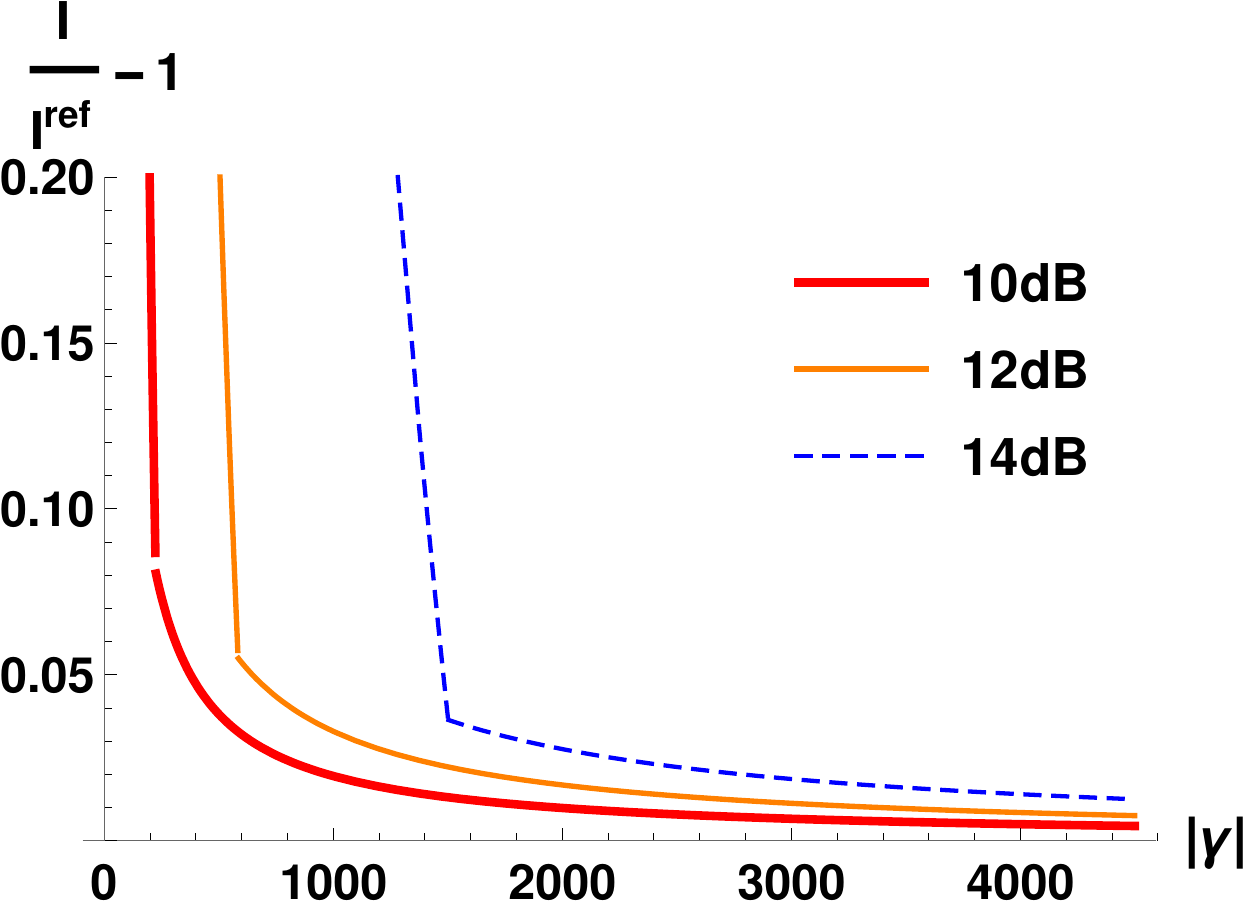}\end{center}  &  \begin{center}\includegraphics[width=.25\textwidth]{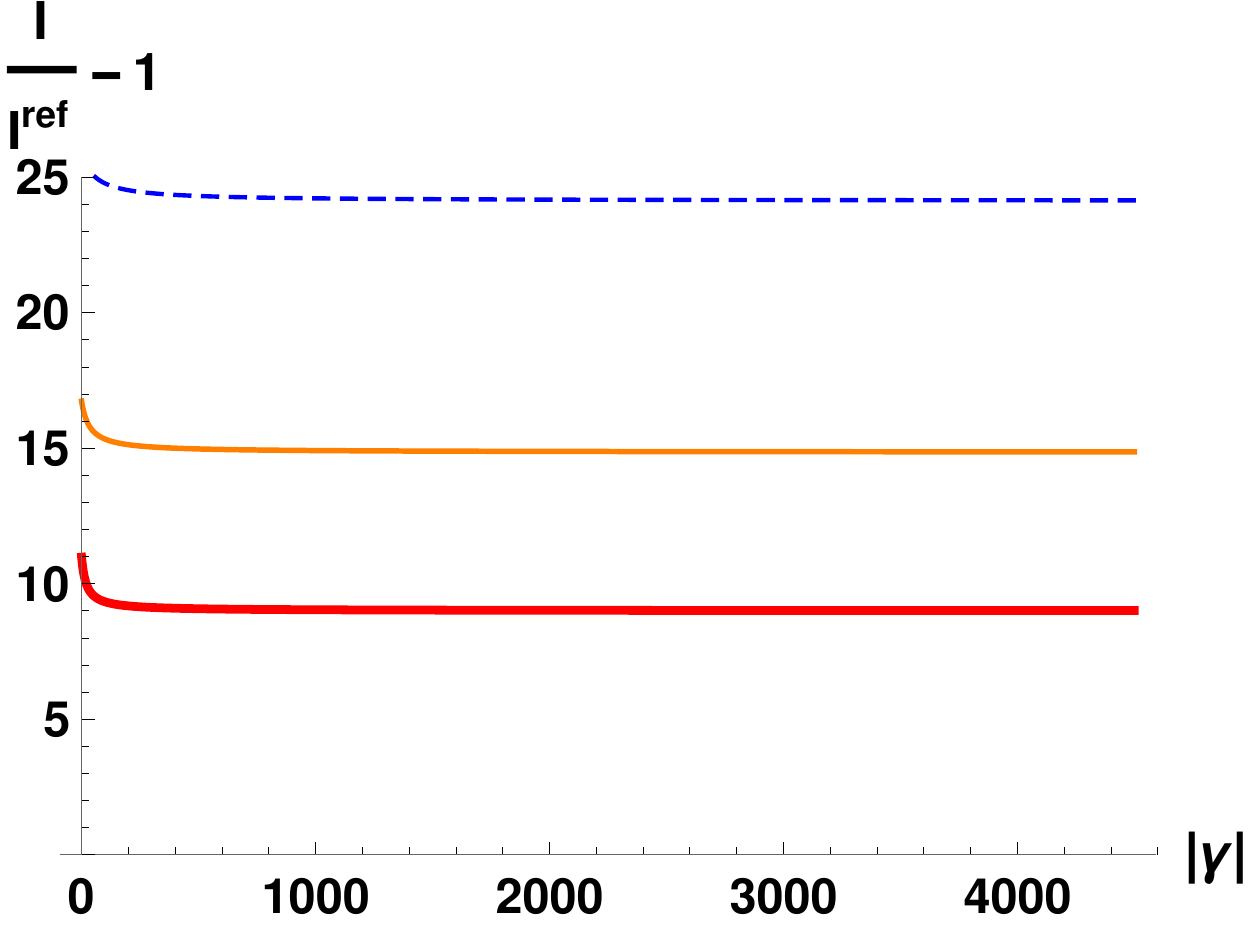}\end{center} \\
	 \begin{center}\includegraphics[width=.25\textwidth]{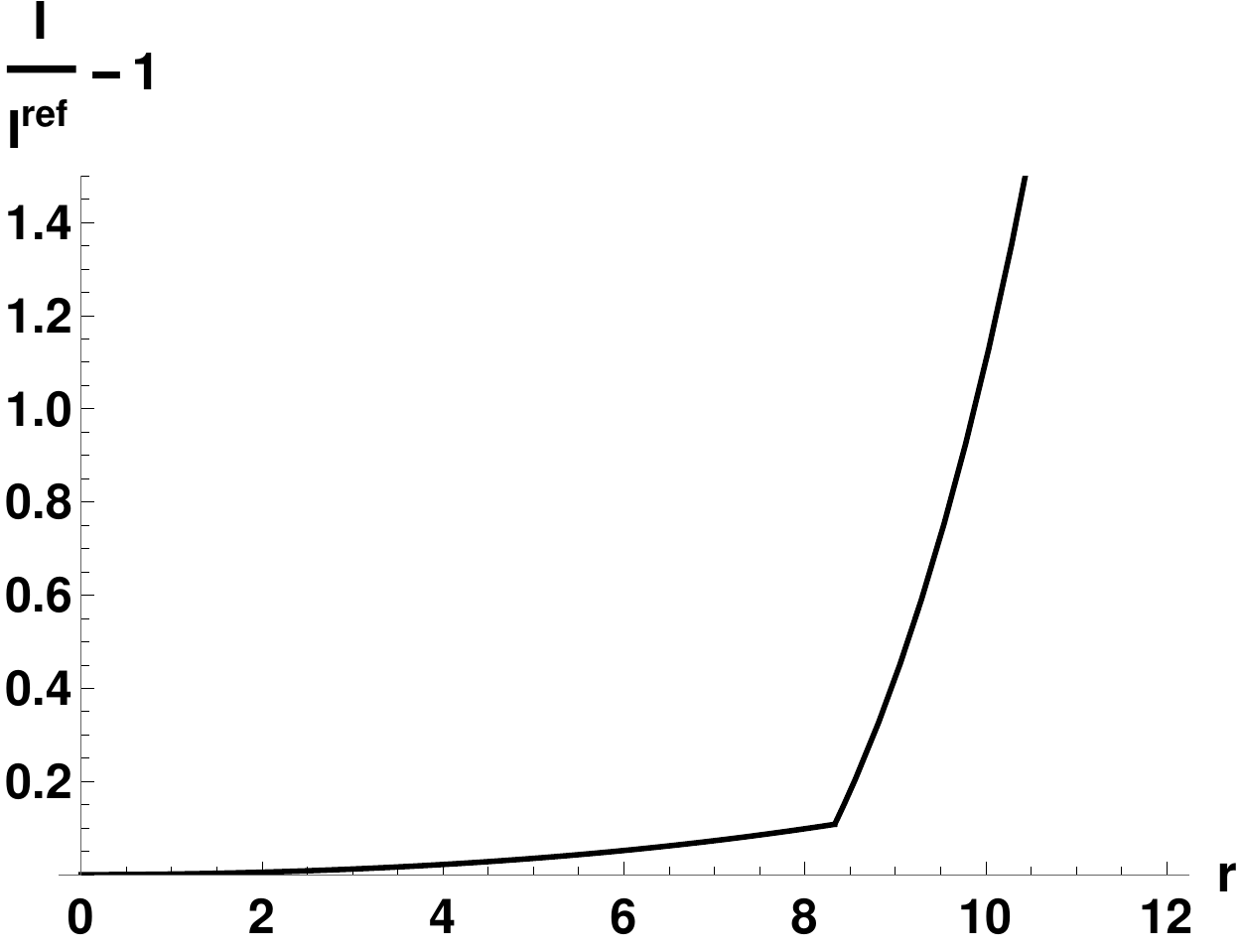}\end{center}  &  \begin{center}\includegraphics[width=.25\textwidth]{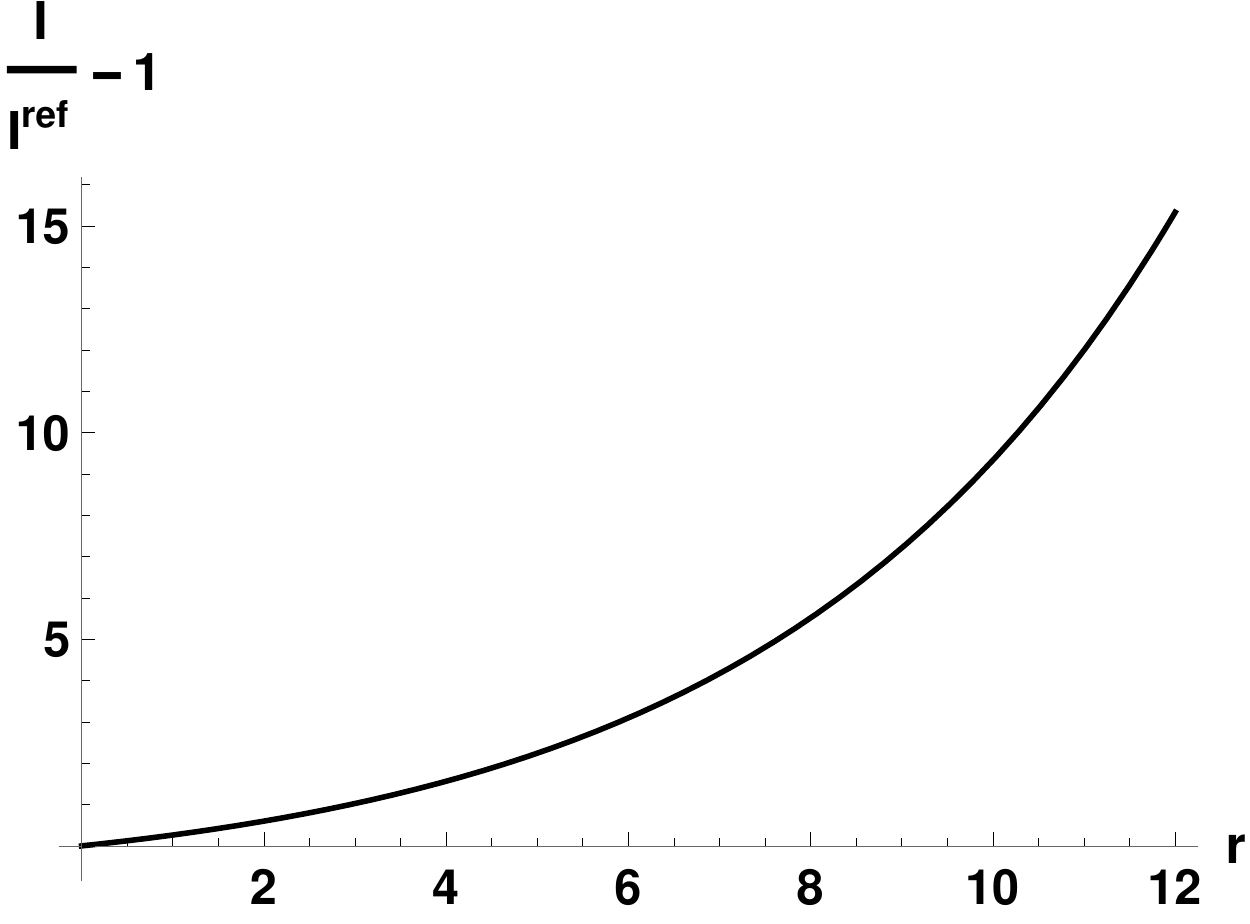}\end{center} \\
	 \begin{center}\includegraphics[width=.25\textwidth]{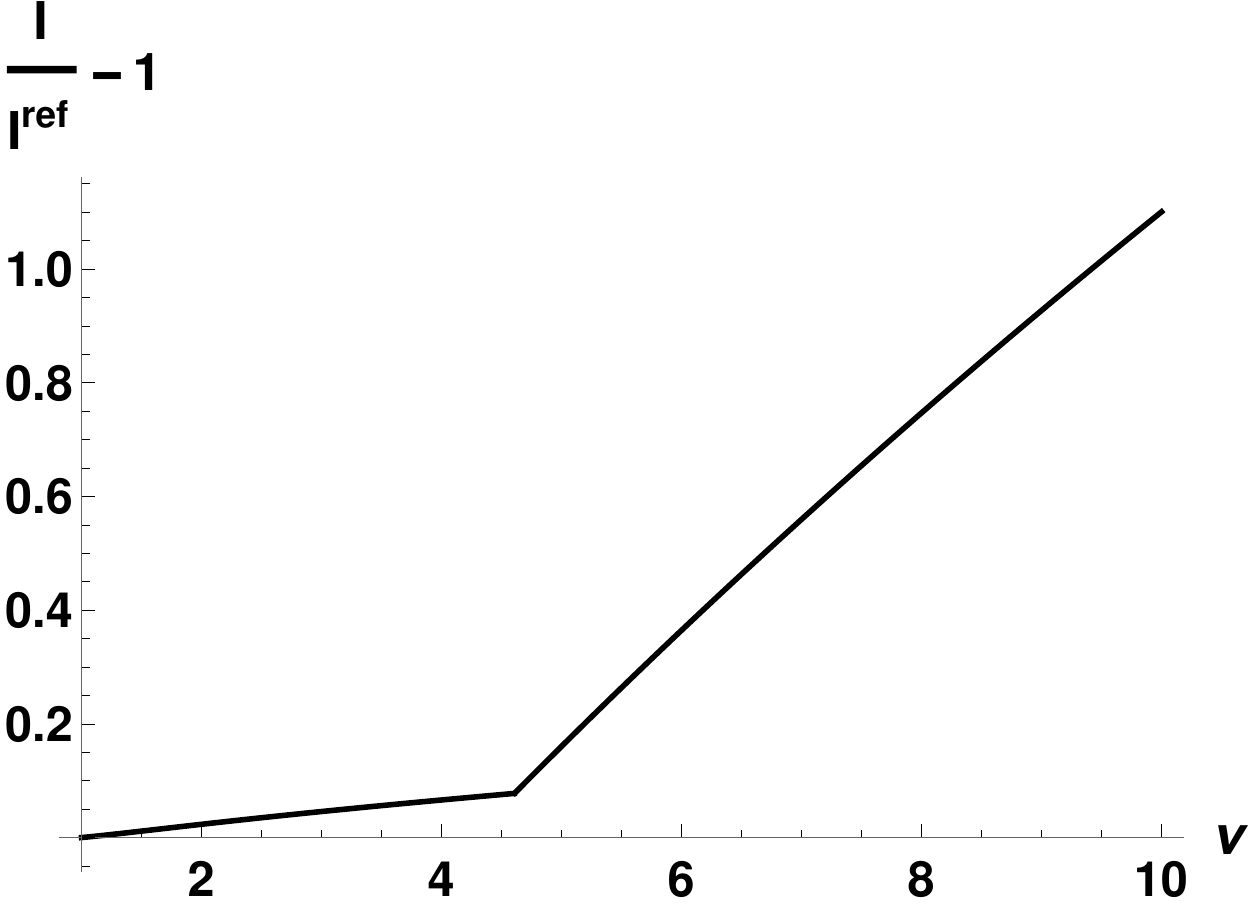}\end{center} &  \begin{center}\includegraphics[width=.25\textwidth]{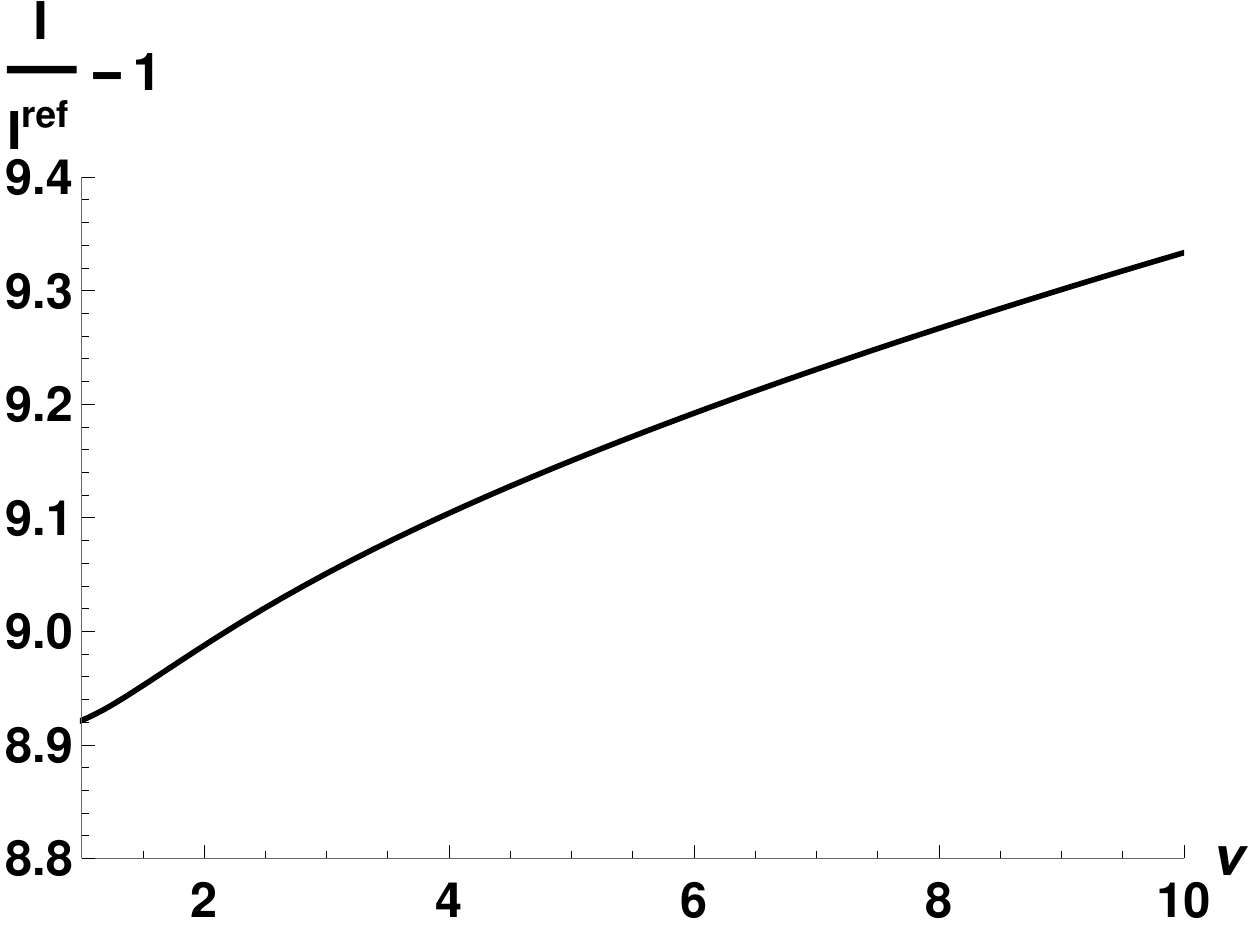}\end{center} \\
	 \hline
	\end{tabular}
	 \caption{Left column: $\phi_1+\phi_{d}=0$, right: $\phi_1+\phi_{d}=\frac{\pi}{2}$. The top figures are schematic depictions of Wigner function of the states in both modes. The curves below show the evolution of $\frac{I_F^{\text{opt}}-I_F^{\text{ref}}}{I_F^{\text{ref}}}$ as a function of the other parameters. From top to bottom: evolution with $\lvert\gamma\rvert$, for$\nu=10$ and various squeezing; evolution with $r$ (in dB), for $\lvert\gamma\rvert=100$ and $\nu=10$; evolution with $\nu$, for $\lvert\gamma\rvert=100$ and $r = 10 dB$ of squeezing. In the case $\phi_1+\phi_{d}=0$, one can distinguish two distinct regimes; the optimal strategy changes when one goes from one regime to the other.}
	 \label{courbes}
\end{figure}

\section{Discussion}
\label{secdiscussion} 

In this section, we are going to discuss possible implications of our findings and how they could be linked to other open research questions. We will first see how our work could be interpreted in a \textit{resource-theoretic} framework. Then we will discuss other possible source of advantage, and how they could be linked to squeezing.

\subsection{Resource theories}

The ability to perform tasks intractable by classical means can be used to define and quantify nonclassicality. This is best captured by resource theories, an emerging formalism that gives a precise meaning to the intuitive notion of resource. Such a theory identifies \textit{free} (or \textit{resourceless}) states, together with \textit{free} operations that cannot create or increase the resource in question. In general, these operations include free unitaries, and more generally quantum maps obtained by adding free ancillas, applying free unitaries on the whole system, then tracing out the ancillas. For instance, for entanglement, free states are separable states and free operations are local operations and classical communication (LOCC). 

 A \textit{resource monotone} $Q$ is then defined to quantify the resource in an arbitrary state. To be a valid quantifier, this function must verify a few important properties. In particular, it must be convex, non-increasing under free operations, and vanish for (and only for) free states. An additional condition, \textit{strong monotonicity}, \textit{i.e.} monotonicity under selective operation on average, is often considered \cite{baumgratz_quantifying_2014}.

 This formalism has already been used to describe various quantities such as coherence \cite{baumgratz_quantifying_2014,streltsov_colloquium:_2017}, asymmetry \cite{ahmadi_wignerarakiyanase_2013}, non-Gaussianity \cite{takagi_convex_2018,albarelli_resource_2018}, or athermality and the laws of thermodynamics \cite{brandao_second_2015,lostaglio_description_2015}. In recent works \cite{kwon_nonclassicality_2018,yadin_operational_2018}, \textit{quadrature displacement} estimation was studied in a resource-theoretic perspective. It was found that the performance of a state for such estimation could be used to quantify the nonclassicality of the state. Now the question is: can advantage in \textit{phase} estimation protocol be similarly used as nonclassicality measure? To make a step in this direction, we will now seek to interpret our metrological advantage $\mathcal{A_G}$ in a resource-theoretic setting.\\

 The definition of quantum metrological advantage provided in Eq.(\ref{metroadvFTQL}) is only valid for Gaussian states. To make a resource-theoretic analysis, it is desirable to extend it to mixtures of Gaussian states. It would also allow to take into account preparation imperfections other that the thermal noise already considered. The difficulty here is that a mixture of Gaussian states is not Gaussian in general; therefore there is no longer a well-defined notion of temperature. It is unclear what the reference state should be if the probe is, \textit{e.g.}, a mixing of two Gaussian states with different temperatures. A possible solution is to perform convex roof minimization, by defining:
\begin{equation}
	\mathcal{A}(\rop) = \text{Min}_{\{p_j,\rop_j\} }\Big( \sum_j p_j \mathcal{A}_G(\rop_j) , 0 \Big)
\end{equation}
Where $\rop=\sum_j p_j \rop_j$, and $\rop_j$ are isotropic Gaussian states. Here we are only performing minimization over Gaussian decomposition; that is, if a state can be written as a mixture of Gaussian states in several different ways, we optimize only over these expressions. Note that $\mathcal{A}$ can be extremely challenging to compute; to the best of our knowledge, it is not even known whether the decomposition of a state into Gaussian component is unique or not. Note also that not all states can be expressed as the sum of Gaussian states. Still, $\mathcal{A}$ yields interesting insight on phase estimation advantage as a resource theory, as we will discuss now.\\

$\mathcal{A}$  has the following properties:

\begin{enumerate}
	\item $\mathcal{A}(\rop)=0$ if and only if $\rop$ is a (convex combination of) displaced thermal state,
	\item $\mathcal{A}$ is non-increasing (actually invariant) under PLO $\Uop$,
	\item $\mathcal{A}$ is convex: $\mathcal{A}$$\Big(\sum_i p_i \rop_i\Big)\leq \sum_i p_i \mathcal{A}(\rop_j)$
\end{enumerate}

Clearly, $\mathcal{A}$ possesses several properties of a resource monotone. However, our framework possesses several properties that are not so easily captured by standard resource theory. In particular, we found two operations that are traditionally considered free, but which turn out to increase $\mathcal{A}$.

First, consider the following protocol: add a coherent state in a third mode, perform PLOs on the whole system, then trace out the third mode. In resource-theoretic terms, this corresponds to an interaction with a free ancilla. Normally, this is a free operation, under which the resource monotone is non-increasing. However, in Appendix \ref{appendixF}, we argue that this protocol can actually increase $\mathcal{A_G}$. More specifically, this protocol results in a displacement of the state, which in turns enhance both $I_F^{\text{opt}}$ and $I_F^{\text{ref}}$. In some cases, this leads to a net increase of $I_F^{\text{opt}}-I_F^{\text{ref}}$. Interestingly, this issue does not arise for the phase estimation protocols studied in \cite{kwon_nonclassicality_2018} and \cite{yadin_operational_2018}. This is because phase estimation precision can be increased by displacing the input state, but displacement estimation can not (a pure coherent state and the vacuum, for instance, have both a QFI of $2$ for displacement estimation). 

Second, as we have already discussed, $\mathcal{A}$ can increase with $\nu$; hence, with our definitions, increasing the thermal noise is not a free operation either.

A possible way to circumvent these issues would be to renormalize the advantage by the FTQL in some way. However, in Appendix \ref{appendixG}, we show that direct renormalization is not sufficient to do this.

\subsection{Link between metrological advantage and other non-classical features}

So far, we have studied the link between metrological advantage and squeezing for Gaussian states. To conclude our work, we will now briefly discuss how other non-classical features may lead to advantage, and the relation between these quantities and squeezing.

The paradigmatic example of non-classical resource is entanglement. When trying to characterize metrologically useful states, it is natural to assume that there is a link between entanglement and metrological advantage.
Quick examination, however, reveals that the situation is more complex. First of all, there are states which are clearly entangled, yet fail to achieve sub-shot-noise scaling. In \cite{hyllus_not_2010}, Hyllus and \textit{al.}. have studied extensively pure states with fixed number of bosonic probes, and exhibited an entire family of entangled state that do not achieve metrological advantage. 

Second, there are states that achieve advantage yet seemingly contain no entanglement at all: a simple example is a one-mode displaced squeezed state, combined with homodyne detection.

Third, there is another, more conceptual, issue. PLOs can create entanglement in the system: a well-known example is the Hong-Ou Mandel protocol. But given their experimental availability, PLOs should be considered as free operations. Since free operations cannot create resource, this mean entanglement can not be a resource for metrology.\\

Lately, several authors (for instance \cite{demkowicz-dobrzanski_chapter_2015} and \cite{sahota_physical_2016}) have shown this third issue can be circumvented by focusing on \textit{particle} entanglement rather that \textit{mode} entanglement. Mode entanglement describes correlations in a Fock basis, while particle entanglement describes correlations between individual photons. Consider, for instance, the state with one photon in each mode. In the Fock basis $\ket{n_1}\ket{n_2}$ (with $n_i$ meaning $n$ photons in mode $i$), this state reads $\ket{1}\ket{1}$ and is clearly separable. In the individual photon basis, however, the symmetrization postulate imposes to write $\frac{\ket{\psi^a_1}\ket{\psi^b_2}+\ket{\psi^b_1}\ket{\psi^a_2}}{\sqrt{2}}$, where $\ket{\psi^{a(b)}_{i}}$ means the first (second) photon is in the mode $i$. Hence, this state is particle-entangled.

On the particle basis, PLOs act on each particle individually. Therefore, these operations cannot create particle-entanglement, but only "reveal" particle entanglement by turning it into mode entanglement; this is what happens in the Hong-Ou Mandel protocol. However, the amount of particle entanglement remains unchanged \footnote{This remark may suggest that particle entanglement is a necessary condition for mode entanglement: this is, however, not correct. One can find mode-entangled states that are particle-separable; for instance, \unexpanded{$\frac{1}{2}(\ket{\psi^a}+\ket{\psi^b})_1(\ket{\psi^a}+\ket{\psi^b})_2=\frac{1}{2}(\ket{2}^a\ket{0}^b + \ket{0}^a\ket{2}^b + \sqrt{2}\ket{1}^a\ket{1}^b)$}.}. This connexion between mode and particle entanglement has been studied by several authors, see \textit{i.e.} \cite{killoran_extracting_2014}. \\

However, considering particle entanglement does not solve the first two issues mentioned earlier. For instance, the states studied by Hyllus \textit{and al.}. \cite{hyllus_not_2010} \textit{are} particle-entangled, and still fail to achieve metrological advantage. The physical nature of particle-entanglement and its usefulness for metrology is still the subject of debates in the literature. In \cite{braun_quantum-enhanced_2018}, for instance, the authors challenge the link between particle entanglement and metrology, and review other possible sources for advantage. (Note though that the definition of particle entanglement in this paper is not exactly the same as in \cite{demkowicz-dobrzanski_chapter_2015} which we have discussed).\\

We will now argue that particle-entanglement may be a sufficient condition for metrological advantage in the case of non-pure Gaussian states. We formulate the following conjecture:

\textbf{Conjecture 1}
\textit{If $\rop$ is a two-mode displaced thermal state, then $\rop$ is particle-separable.}

We already now Conjecture 1 is true for pure states. To the best of our knowledge, however, no one has extended this to mixed states. If this conjecture is true, then, using Theorem 1, we can show that particle-entanglement implies metrological advantage, when one considers bimodal isotropic Gaussian states. The condition of Gaussianity excludes the states considered in \cite{hyllus_not_2010}; thus this conjecture is consistent with previously known results. In Appendix \ref{separability displtherm}, we show that in the two-photon subspace, bimodal displaced thermal state are separable. This supports the conjecture; proving it, however, would require a full analysis of displaced thermal states in the first-quantization basis. \\

Even if Conjecture 1 is true, however, we still need to account for the performances of single-mode displaced squeezed state, which of course can be neither mode- nor particle-entangled. A possible origin for the metrological usefulness of these states is the superposition between different particle \textit{numbers}. (In other words, we have to consider not only entanglement within the subspaces corresponding to fixed $\moy{\Nop}$, but also coherences between those subspaces).

\begin{figure}[H]
\begin{center}
\includegraphics[scale=0.17]{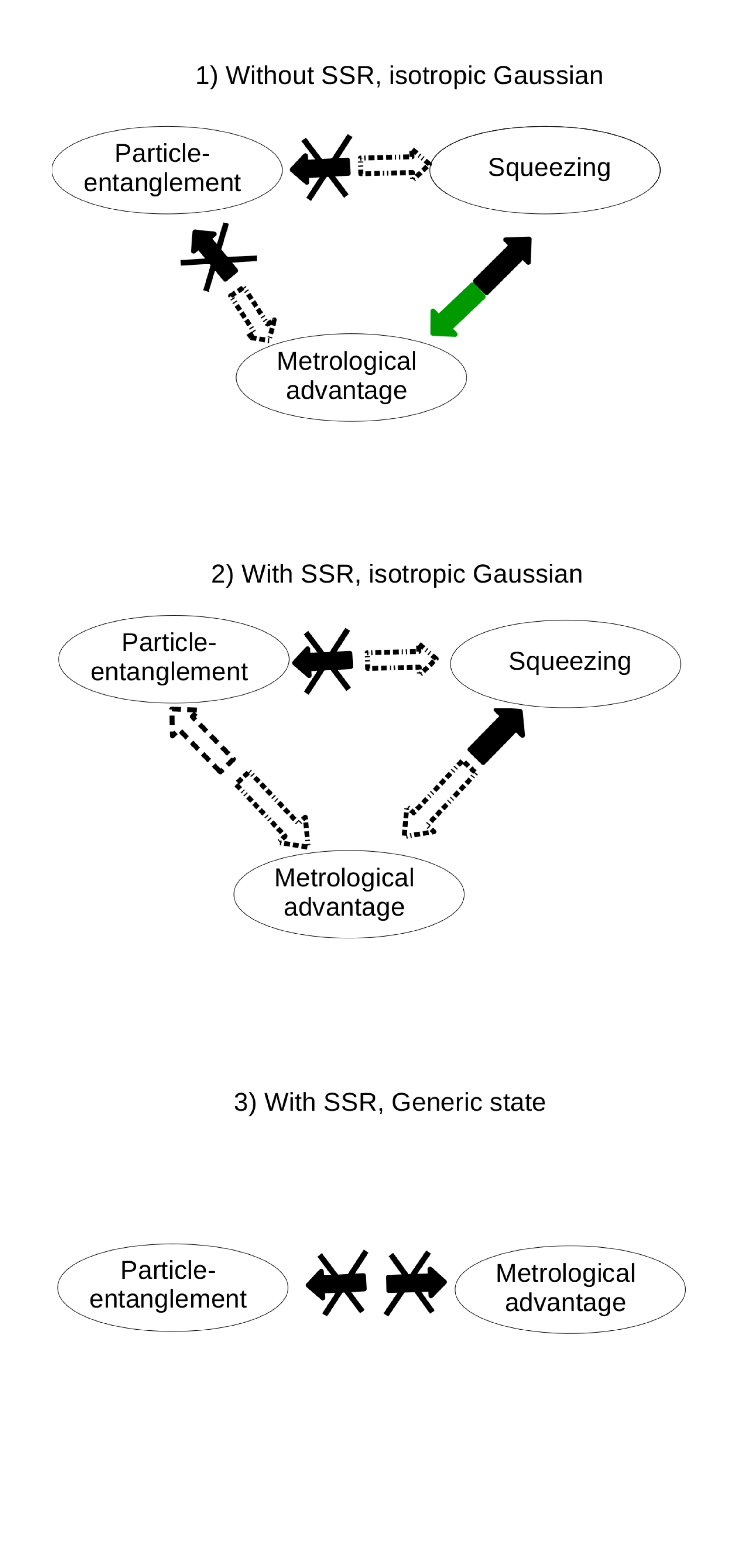}
\end{center}
\caption{Summary of the links between squeezing, particle entanglement, and metrological advantage, with and without SSR, for Gaussian or arbitrary states. Full arrows are established results, dotted arrows are conjectures. Our results correspond to the green arrow on the first cartoon. The third cartoon corresponds to the results of \cite{hyllus_not_2010}. For non-Gaussian states without SSR, no general results are known to the best of our knowledge.}
\label{summary}	
\end{figure}

A simple way to study the role of number superposition is to see what happen when it is suppressed. For massive particles such as atoms, number superposition are naturally suppressed by superselection rules (SSR). For photons we can mimic it by applying restriction at the measurement level. So far we have authorized the measure of all observables, including homodyne measurement. If we have only access to observables that commute with the total number of photons (for instance, if we have access to photon-counter but no perfect phase reference), then this is equivalent to suppressing the number superposition in the input state. This problem can be expressed in the \textit{quantum reference frame} language \cite{bartlett_reference_2007}. Metrological protocols with such limitations have already been studied in specific cases, in particular for pure input state \cite{jarzyna_quantum_2012,safranek_quantum_2015}. It would be quite interesting to see whether Theorem 1, or a modified version thereof, still holds in that case. In \figref{summary}, we have summarized the known results, and the conjectures we just formulated. We believe further understanding of the link between particle entanglement, number superposition, squeezing and metrological advantage will be important to define phase estimation-based nonclassicality quantifier that can be formulated in a resource-theory framework.

\section{Conclusion}
\label{secconclusion}

To better understand the link between nonclassicality and metrological advantage, we have introduced a definition of metrological advantage which takes thermal noise into account. For bimodal non-pure isotropic Gaussian states, we have demonstrated that squeezing is a necessary and sufficient condition to achieve metrological advantage. We have discussed the properties of our measure, and the challenges that remain to be overcome to achieve full-fledged resource theory of phase estimation. We hope our work will be a step towards using phase estimation as a nonclassicality quantifier. \\

This work focused on the role of squeezing in phase estimation. To go further, it would be interesting to study in more details the role of other quantities such as particle entanglement and number superposition. We have briefly reviewed some challenges and open questions concerning these issues. We may also generalize our study by considering multi-modes anisotropic states, or states that cannot be decomposed as a sum of Gaussian states. The key question then would be how to properly define a reference state, since in this case we cannot match a displaced thermal state with each state. Finally, we may also try to include noise at the evolution level, in addition to the preparation level.

\vspace{20 cm}

\section*{Acknowledgments} 
We acknowledge useful discussions with N.Treps. S.F. acknowledges support from the French Agence Nationale de la Recherche (SemiQuantRoom, Project No. ANR14-CE26-0029) and from the PRESTIGE program, under the Marie Curie Actions- COFUND of the FP7.

\bibliographystyle{apsrev4-1}
\bibliography{References_pra}

\appendix

 \section{Gaussian states in phase space}
 \label{appendixA}

In the following, it will be very useful to work with a phase-space representation of our states, instead of the usual Hilbert space representation. We will make extensive use of the parameterization and expression developed in \cite{safranek_gaussian_2016} and \cite{safranek_optimal_2016}.
 For a $q$-mode state, we can collect the creation and annihilation operators in a vector $\mathbf{A}=(\hat{\textbf{a}},\hat{\textbf{a}}^{\dagger})^T$, with $\hat{\textbf{a}}=(\hat{a}_1,...,\hat{a}_q$). We will study the symmetric characteristic function $tr[\rop e^{\mathbf{A}^{\dagger} K \bm{\xi}}]$, where $\bm{\xi}$ is in $\mathbb{C}^{2q}$ and can be written like $\bm{\xi}=\mathbf{x}\oplus\bar{\mathbf{x}}$, and $K$ is the so-called symplectic form, which here reads as $K=\left[ {\begin{array}{cc}
\openone_q & 0\\
0 & -\openone_q\\	
\end{array}} \right]$
where $\openone_q$ is the $q\times q$ identity matrix. For Gaussian states, the characteristic function takes a Gaussian form:
\begin{equation}
tr[\rop e^{\mathbf{A}^{\dagger} K \bm{\xi}}]=\exp\Big(-\frac{\xivect^{\dagger}\sigma\xivect}{4}-i\ddagvect K\xivect\Big)
\end{equation}

Here $\sigma$ is the covariance matrix and $\dvect$ the displacement vector, defined as:
\begin{subequations}

\begin{equation}
	\dvect^i=tr[\rop\mathbf{A}^i]
\end{equation}
\begin{equation}
	\sigma^{ij}=tr[\rop\{\mathbf{A}^i-\dvect^i,\mathbf{A}^{j\dagger}-\dvect^{j\dagger}\}]	
\end{equation}
\end{subequations}

For a $q$-mode thermal state, one has $\dvect=0$ and $\sigma=diag(\nu_1,...\nu_q,\nu_1,...,\nu_q)$, where $\nu_i=\text{cotanh}(\frac{\hbar\omega_i}{2kT_i})$.

In phase space, phase-shifting, beam-splitter, and squeezing unitaries act as symplectic operations $\Sigma$, \textit{i.e.}, $2q\times 2q$ matrices that verify $\Sigma K \Sigma^{\dagger}=K$. These matrices define a complex representation of the \textit{real symplectic group} Sp(2q,$\mathbb{R}$). Under such unitaries, first and second moments transform as: $\dvect\rightarrow \Sigma \dvect$ and $\sigma\rightarrow \Sigma \sigma \Sigma^{\dagger}$. Displacement operators $\Dop(\bm{\gamma})$ act as $\dvect\rightarrow\dvect+\bm{\gamma}$.

Now, let us consider the decomposition introduced in Eq.\eqref{symplecform} and \eqref{symplecdef}. If we consider these equations in phase space, we arrive at the following:
\begin{subequations}
\label{dsigma}
\begin{equation}
	\dvect=(\bm{\gamma},\overline{\bm{\gamma}})^T
\end{equation}
\begin{widetext}
\begin{eqnarray}
	 \nonumber \sigma & = & \Rmat_1(\phi_1)\Rmat_2(\phi_2)\Bmat(\theta)\Rmat_{as}(\psi)\Smat_1(r_1)\Smat_2(r_2)\Rmat_{as}(\psi_1) \Bmat(\theta_1)diag(\nu_1,\nu_2,\nu_1,\nu_2)(.)^{\dagger}\\
	 & = & \mathcal{M} diag(\nu_1,\nu_2,\nu_1,\nu_2) \mathcal{M}^{\dagger}
\end{eqnarray}
\end{widetext}
\end{subequations}
Where $diag$ denotes diagonal matrix. These equations involve the representation of symplectic operations in phase space:\\

$\Rmat_1(\phi_1)=\begin{bmatrix}
e^{-i\phi_1} & 0 & 0 & 0 \\
0 & 1 & 0 & 0 \\
0 & 0 & e^{i\phi_1} & 0 \\
0 & 0 & 0 & 1\\
\end{bmatrix} $ \\

$\Rmat_2(\phi_2)=\begin{bmatrix}
1 & 0 & 0 & 0 \\
0 & e^{-i\phi_2} & 0 & 0 \\
0 & 0 & 1 & 0 \\
0 & 0 & 0 & e^{i\phi_2} \\
\end{bmatrix}$\vspace{10pt}

$\Bmat(\theta)=\begin{bmatrix}
\cos(\theta) & \sin(\theta) & 0 & 0 \\
-\sin(\theta) & \cos(\theta) & 0 & 0 \\
0 & 0 & \cos(\theta) & \sin(\theta) \\
0 & 0 & -\sin(\theta) & \cos(\theta) \\
\end{bmatrix}$\vspace{10pt}

$\Smat_1(r_1)= \begin{bmatrix}
\cosh(r_1) & 0 & -\sinh(r_1) & 0 \\
0 & 0 & 0 & 0 \\
-\sinh(r_1) & 0 & \cosh(r_1) & 0 \\
0 & 0 & 0 & 0 \\
\end{bmatrix}$\vspace{10pt}

$\Smat_2(r_2)= \begin{bmatrix}
0 & 0 & 0 & 0 \\
0 & \cosh(r_2) & 0 & -\sinh(r_2) \\
0 & 0 & 0 & 0 \\
0 & -\sinh(r_2) & 0 & \cosh(r_2) \\
\end{bmatrix}$\hspace{40pt}

\vspace{10pt}

 $\Rmat_{as}(\phi)=\Rmat_1(\phi)\Rmat_2(-\phi)$ \\

As of the symplectic eigenvalues $\nu_i$, they can be obtained by solving the usual eigenvalue problem for the matrix $A=K\sigma$. Since $K$ is invertible, the $\nu_i$ are uniquely define for a given Gaussian state. \\

Note that $\Bmat^{\dagger}=\Bmat^{-1}$ and $\Rmat_i^{\dagger}=\Rmat_i^{-1}$; thus, for isotropic states $\nu_1=\nu_2=\nu$, we can simplify Eq.(\ref{dsigma}) and obtain:
\begin{equation}
\label{decomposition}
	\sigma=\nu\Rmat_1(\phi_1)\Rmat_2(\phi_2)\Bmat(\theta)\Rmat_{as}(\psi)\Smat_1(r_1)\Smat_2(r_2)(.)^{\dagger}
\end{equation}

Finally, let us give the complete expression for the mean number of photons:

\begin{equation}
\label{Nb}
	\langle\Nop\rangle=\adag\aop+\bdag\bop = \frac{1}{4}tr[\sigma]-1+\lvert \bm{\gamma}\rvert^2
\end{equation}
(Note that the modulus squared of the displacement vector, $\lvert \bm{d}\rvert^2$, is equal to $2(\lvert \bm{\gamma}\rvert^2)$. 
Due to the unitarity of $\Rmat_1$, $\Rmat_2$ and $\Bmat$, the phase shifting and mode-mixing operation preserve the trace of the covariance matrix. Therefore, we have:

\begin{eqnarray*}
tr[\sigma]& = & tr[\nu\Smat_1(r_1)\Smat_2(r_2)\Smat_2(r_2)^{\dagger}\Smat_1(r_1)^{\dagger}]\\
 & = & tr[\nu\Smat_1(2r_1)\Smat_2(2r_2)]
\end{eqnarray*}

Which yields:
\begin{eqnarray}
	tr[\sigma] & = & 2\nu(\cosh{2r_1}+\cosh{2r_2})\\
	 & = & 4\nu+4\nu(\sinh(r_1)^2+\sinh(r_2)^2) \nonumber
\end{eqnarray}

\section{Expression of the metrological advantage}
\label{appendixB}

For a Gaussian state, it is possible to extract exact general formulas for the QFI, see for instance \cite{safranek_optimal_2016,safranek_gaussian_2016,monras_phase_2013,jiang_quantum_2014}. In particular, in \cite{safranek_gaussian_2016} and \cite{safranek_optimal_2016}, the expression for two-mode Gaussian state evolving under the channel $\rop \rightarrow e^{-ix\Jy}\rop e^{ix\Jy}$ was fully derived. We will use these expressions adapted to our case. With all the notations introduced, the QFI reads: 

\begin{eqnarray}
\label{eqpqfi}
	I_F(\rop_{x}^{\uvect}) = 4m^2\frac{\nu^2}{\nu^2+1}(\sinh(2r_1)^2+\sinh(2r_2)^2) \nonumber \\ 
	+\frac{8\nu^2}{\nu^2+1} \Big(p^2\sinh(r_1-r_2)^2 +o^2\sinh(r_1+r_2)^2\Big) \nonumber \\
	+\frac{4\lvert \gamma\rvert^2}{\nu}\Big(  e^{2r_1} \kappa^2 + e^{-2r_1} \delta^2 + e^{2r_2} \upsilon^2 + e^{-2r_2} \lambda^2\Big)
\end{eqnarray}
Where we have defined 
\begin{align}
\label{defpara}
\nonumber
m=\sin(2\theta)\sin(\phi_1-\phi_2) \\ \nonumber
o=\cos(2\theta)\sin(\phi_1-\phi_2)\cos(2\psi)+\cos(\phi_1-\phi_2)\sin(2\psi) \\ \nonumber
p=\cos(2\theta)\sin(\phi_1-\phi_2)\sin(2\psi)-\cos(\phi_1-\phi_2)\cos(2\psi) \\ \nonumber
\kappa=\cos(\alpha) \sin(\theta)\cos(\tilde \phi_2+\psi) + \sin(\alpha)\cos(\theta)\cos(\tilde\phi_1+\psi) \\ \nonumber
\delta=\cos(\alpha) \sin(\theta)\sin(\tilde \phi_2+\psi)+\sin(\alpha)\cos(\theta)\sin(\tilde\phi_1+\psi) \\ \nonumber
\upsilon=\cos(\alpha) \cos(\theta)\cos(\tilde\phi_2-\psi) - \sin(\alpha)\sin(\theta)\cos(\tilde\phi_1-\psi) \\ 
\lambda=\cos(\alpha) \cos(\theta)\sin(\tilde\phi_2-\psi) - \sin(\alpha)\sin(\theta)\sin(\tilde\phi_1-\psi)
\end{align}

with $\tilde\phi_1=\phi_1+\phi_{d2}$ and $\tilde\phi_2=\phi_2+\phi_{d1}$.

 Notice the following properties:
$m^2+p^2+o^2=1$ and $\kappa^2+\delta^2+\upsilon^2+\lambda^2=1$.

Now, let us express the precision reference. Using (\ref{Nb}), we may rewrite it as:
\begin{equation}
	4\frac{\langle \Nop\rangle +1}{\nu}-4=\frac{tr[\sigma]}{\nu}+\frac{4\lvert\gamma\rvert^2}{\nu} - 4 = 4(\sinh(r_1)^2+\sinh(r_2)^2)+\frac{4\lvert\gamma\rvert^2}{\nu}	
\end{equation}

And we obtain after a little rewriting:
\begin{widetext}
\begin{equation}
	\label{eqdep}
	I_F-I_F^{\text{ref}} = I_F(\rop_{x}^{\uvect})-\Big(4\frac{\langle \Nop\rangle+1}{\nu}-4\Big)\\
	  =2[m^2Z + o^2 X +p^2Y] + \frac{4\lvert\gamma\rvert^2}{\nu}\Big(e^{2r_1}\kappa^2 + e^{-2r_1}\delta^2 + e^{2r_2}\upsilon^2 + e^{-2r_2}\lambda^2 -1 \Big)
\end{equation}
\end{widetext}

Where we have defined:
\begin{subequations}
\begin{equation*}
 Z=2 \frac{\nu^2}{\nu^2+1}\Big(\sinh^2(2r_1)+\sinh^2(2r_2)\Big) - 2\big(\sinh^2(r_1)+\sinh^2(r_2)\big) 
 \end{equation*}
 \begin{equation*}
  Y=4 \frac{\nu^2}{\nu^2+1}\Big( \sinh^2(r_1-r_2) \Big) - 2\big(\sinh^2(r_1)+\sinh^2(r_2)\big)	
 \end{equation*}
 \begin{equation*}
 X=4 \frac{\nu^2}{\nu^2+1}\Big( \sinh^2(r_1+r_2) \Big) - 2\big(\sinh^2(r_1)+\sinh^2(r_2)\big)	
 \end{equation*}
 \end{subequations}

 \section{Proof of Theorem 1}
 \label{appendixC}

Our goal is to prove that, by using PLOs, we can always obtain a set of parameters such that Eq.(\ref{eqdep}) becomes strictly positive.

First, we need to understand how the application of a PLO operator $\Uop$ allows us to control the various parameters that describe $\rop$. $\Uop$ may be decomposed using Euler's angles: 
\begin{equation}
\label{Udecompo}
\Uop=\Rop_{as}(a)\Bop(b)\Rop_{as}(c)
\end{equation}
with $a$, $b$ and $c$ arbitrary angles. To find the optimal precision, we may directly write down the state $\rop^U$, compute its metrological advantage, then use numerical methods to optimize this expression with respect to $a$, $b$ and $c$. 

Here, however, we will adopt a slightly more specific strategy, that will allow for analytical treatment, while remaining general enough to prove our main result.

Let us consider a state described by $(\nu,\lvert\gamma\rvert,\alpha^0,\phi_{d1}^0,\phi_{d2}^0,\phi_1^0,\phi_2^0,\theta^0,\psi^0,r_1,r_2)$. Without loss of generality, we can set $r_1\geq0$ and $r_2\geq0$ (any sign difference can be absorbed in the other parameters). If we apply the operator $\Uop=\Rop_{as}(-\psi^0 + \frac{\pi}{4})\Bop(-\theta^0)\Rop_{as}(\frac{\phi_2^0-\phi_1^0}{2})$, we obtain: 

\begin{widetext}
\begin{eqnarray*}
\Rmat_{as}(-\psi^0 + \frac{\pi}{4})\Bmat(-\theta^0)\Rmat_{as}(\frac{\phi_2^0-\phi_1^0}{2}) \Big(\Rmat_1(\phi_1^0)\Rmat_2(\phi_2^0)\Bmat(\theta^0)\Rmat_{as}(\psi^0)\Big) & = & \Rmat_1(\frac{\phi_1^0+\phi_2^0}{2})\Rmat_2(\frac{\phi_1^0+\phi_2^0}{2})\Rmat_{as}(\frac{\pi}{4})\\
 & = & \Rmat_1(\frac{\phi_1^0+\phi_2^0}{2} + \frac{\pi}{4})\Rmat_2(\frac{\phi_1^0+\phi_2^0}{2} - \frac{\pi}{4}) 	
\end{eqnarray*}
\end{widetext}

According to the decomposition \eqref{decomposition}, this means we have $\theta=0$, $\psi=0$ and $\phi_1-\phi_2=\frac{\pi}{2}$. The other parameters ($\alpha$, $\phi_{d1}$ and $\phi_{d2}$) are also modified, in a non-trivial way.\\

 Hence, using PLOs, all isotropic Gaussian states can be mapped to some state $(\nu,\lvert\gamma\rvert,\alpha^i,\phi_{d1}^i,\phi_{d2}^i,\phi_1^i=\phi_2^i+\frac{\pi}{2},\theta^i=0,\psi^i=0,r_1,r_2)$, with $r_1\geq0$, $r_2\geq0$, and $\alpha^i$, $\phi_{d1}^i$ and $\phi_{d2}^i$ arbitrary. This means we have $m_i=p_i=0$ and $o_i=1$ in \eqref{defpara}. We can also define the corresponding values $\kappa_i$, $\delta_i$, $\upsilon_i$, $\lambda_i$. From now on, we will exclusively consider these states.\\

 In the proof, we will need the property $Z \geq X \geq \lvert Y \rvert \geq 0$, which can be easily proven using the properties of the sinh function and the fact that $\nu\geq1$. We also recall $m^2+o^2+p^2=\delta^2+\kappa^2+\lambda^2+\upsilon^2=1$. \\
 
Let us now consider Eq.(\ref{eqdep}) with $p=m=0$. It yields: $$\frac{I_F-I_F^{\text{ref}}}{2}=X+\frac{4\lvert\gamma\rvert^2}{\nu}V_i$$
with $V_i=\Big(e^{2r_1}\kappa_i^2 + e^{-2r_1}\delta_i^2 + e^{2r_2}\upsilon_i^2 + e^{-2r_2}\lambda_i^2 -1 \Big)$. Then three cases are possible:

\begin{itemize}
	\item  $\bm{V_i > 0}$: then (\ref{eqdep}) is strictly positive for all $\lvert\gamma\rvert$ and all $\nu$. Note this case is only possible if $r_1\neq0$ or $r_1\neq0$.\\

	\item $\bm{V_i < 0}$: as before, this is only possible if $r_1\neq0$ or $r_1\neq0$. If we keep the same choice of parameters, (\ref{eqdep}) will be positive for small $\lvert\gamma\rvert$, but negative for high $\lvert\gamma\rvert$. However, we can solve this issue by applying an additional phase-shift when the value becomes negative. More precisely, let us define $\lvert\gamma_s\rvert^2=\frac{\nu X}{2\lvert V_i\rvert}$. For $\lvert\gamma\rvert < \lvert\gamma_s\rvert$ (which corresponds either to low displacement or high temperature), our choice of parameters still yields $\frac{I_F-I_F^{\text{ref}}}{2}\geq X + \frac{2\lvert\gamma\rvert^2}{\nu}(V_i)>0$ (since $Z\geq X$). When $\gamma\geq\gamma_s$, we apply the additional phase shift $\Uop=\Rop_{as}(\frac{\pi}{4})$. This sets new values for the parameters: $\phi_1^f=\phi^i_1 + \frac{\pi}{4}$, $\phi_2^f=\phi^i_2 - \frac{\pi}{4}$, $\phi_{d1}^f=\phi_{d1}^i - \frac{\pi}{4}$ and $\phi_{d2}^f=\phi_{d2}^i + \frac{\pi}{4}$. This also gives $\tilde\phi_1^f=\tilde\phi_1^i+\frac{\pi}{2}$ and $\tilde\phi_2^f=\tilde\phi_2^i-\frac{\pi}{2}$ . This, coupled with $\theta^i=0$, yields $\kappa_f=-\delta_i$, $\delta_f=\kappa_i$, $\upsilon_f=\lambda_i$, $\lambda_f=-\upsilon_i$. Of course, the parameter $p$ will not be equal to $0$ anymore (actually, it will even be equal to $1$). However, since $p\leq1$ and $\lvert Y \rvert\leq X$, we have:

\begin{align}
\nonumber
\frac{I_F-I_F^{\text{ref}}}{2} \geq -X + \frac{2\lvert\gamma\rvert^2}{\nu}W_i \geq \frac{2\lvert\gamma\rvert^2}{\nu}(W_i+V_i) \\ \nonumber
W_i =(e^{2r_1}\delta_i^2 + e^{-2r_1}\kappa_i^2 + e^{2r_2}\lambda_i^2 + e^{-2r_2}\upsilon_i^2 -1) 
\end{align}
We have used $-X\geq-\frac{2\lvert\gamma\rvert^2 \lvert V_i\rvert}{\nu}$ (since $\lvert\gamma\rvert\geq\lvert\gamma_s\rvert$), and $V_i\leq0$. Then we have:

\begin{align}
\nonumber
W_i + V_i & = (\kappa_i^2+\delta_i^2)(e^{2r_1}+e^{-2r_1}) + (\upsilon_i^2+\lambda_i^2)(e^{2r_2}+e^{-2r_2}) -2 \\ \nonumber
& > 2(\kappa_i^2+\delta_i^2) + 2(\upsilon_r^2+\lambda_r^2) -2=0
\end{align}
The last inequality is strict when $V_i\neq 0$. Then \eqref{eqdep} is strictly positive.

\item $\bm{V_i=0}$: this is only possible in three cases, namely $r_2=\kappa_i^2=\delta_i^2=0$, $r_1=\delta_i^2=\lambda_i^2=0$, or $r_1=r_2=0$. The latter case is trivial, since it corresponds to displaced thermal states. Let us examine the case $r_2=\kappa_i^2=\delta_i^2=0$ in more details. Coupled with $\theta_i=\psi_i=0$, these conditions imply $\alpha_i=0$. In other words, both displacement and squeezing are zero in the second mode: we have a displaced squeezed state in the first mode and a thermal state in the second. Now, since $r_2=0$, we have $Y=X$. Then we have immediately:

\begin{align}
\nonumber
	\frac{I_F-I^{\text{ref}}_F}{2} & = m^2 Z + (o^2 + p^2) X + \frac{2\lvert\gamma\rvert^2}{\nu}V_i \\ \nonumber
	& = X + m^2 (Z-X) \\ \nonumber
	 & \geq X
\end{align}
\end{itemize}
As long as $\nu>1$ and $r_1\neq0$, then $X>0$, and we have an advantage.

Finally, the case $r_1=\delta_i^2+\lambda_i^2=0$ can be mapped to the previous one by swapping both modes, which concludes the proof.\\

This proof works for non-pure states; however, the cases $V_i>0$ and $V_i<0$ remain the same when $\nu=1$. The next section deals in more details with the case $V_i=0$ (which we shown boils down to $r_2=\theta_i=\alpha_i=\psi_i=0$, \textit{i.e.}, a displaced squeezed state in the first mode and the vacuum in the second.) This analysis was also used to produce the curves in \figref{courbes}.

\section{Specific cases and pure states}
\label{appendixD}

We want to find the optimal precision for an input state $$\rop_i= \Dop(\bm{\gamma})\Rop_1(\phi_1)\Rop_2(\phi_2)\Sop_1(r_1)\rop_{\text{th}}(\nu)(.)^{\dagger}$$ with $\bm{\gamma}=e^{i\phi_d}(1,0)$. Starting from $\rop_i$, we apply an arbitrary PLO $\Uop=\Rop_{\text{as}}(a)\Bop(b)\Rop_{\text{as}}(c)$, and obtain a state $\rop^U$. This state is described by the following set of parameters: $\phi_1^U=a+\frac{\phi_1+\phi_2}{2}$, $\phi_2^U=-a+\frac{\phi_1+\phi_2}{2}$, $\theta^U=b$, $\psi^U=\frac{\phi_1-\phi_2}{2}+c$, $\phi_{d1}^U=\phi_d-c-a$, $\phi_{d2}^U=\phi_d-c+a$, $\alpha^U=-b$. Straightforward calculations then yield: 

\begin{align}
\label{specificase}\nonumber
	I_F(\rop^U)= & \frac{8\nu^2}{\nu^2+1}\sinh{r_1}^2 + \frac{4\lvert\gamma\rvert^2}{\nu} + \\ \nonumber
	& \sin^2(2b)\sin^2(2a)\Big[\frac{4\nu^2}{\nu^2+1}\big(\sinh^2(2r_1)-2\sinh^2(r_1)\big) \\ 
	& + \frac{4\lvert\gamma\rvert^2}{\nu}(e^{2r_1}\sin^2\big(\tilde{\phi})+e^{-2r_1}\cos^2(\tilde{\phi}) - 1\big)\Big]
\end{align}

with $\tilde{\phi}=\phi_d+\phi_1$; this parameter describes the relative orientation of squeezing and displacement. $\tilde{\phi}=0$ means displacement and noise reduction occurs in the same direction. $\tilde{\phi}=\frac{\pi}{2}$ means displacement and squeezing are orthogonal. 

 We need now to optimize \eqref{specificase} with respect to $a$, $b$ and $c$. Three cases are possible:

\begin{itemize}
	\item $V=e^{2r_1}\sin^2\big(\tilde{\phi})+e^{-2r_1}\cos^2(\tilde{\phi}) - 1\geq 0$. Then the optimal protocol is always to set $b=a=\frac{\pi}{4}$. One can show that this amounts only to a simple one-mode phase estimation protocol. 
	We have then: $$I_F^{\text{opt}}= \frac{4\nu^2}{\nu^2+1}\sinh^2(2r_1) + \frac{4\lvert\gamma\rvert^2}{\nu}(V+1)$$

	\item $V<0$, and $-4\frac{\lvert\gamma\rvert^2 V}{\nu} < \frac{4\nu^2}{\nu^2+1}\big(\sinh^2(2r_1)-2\sinh^2(r_1)\big)$. Then the optimal strategy and $I_F^{\text{opt}}$ remain the same. 
	\item $V<0$, and $-4\frac{\lvert\gamma\rvert^2}{\nu}V \geq \frac{4\nu^2}{\nu^2+1}\big(\sinh^2(2r_1)-2\sinh^2(r_1)\big)$. Then the optimal precision is achieved by setting $a=b=0$, that is, sending directly the state in the MZ interferometer. In this case, we have $$I_F^{\text{opt}}=\frac{8\nu^2}{\nu^2+1}\sinh{r_1}^2 + \frac{4\lvert\gamma\rvert^2}{\nu}$$
\end{itemize}

The FTQL is $I_F^{\text{ref}}=4\frac{\lvert\gamma\rvert^2}{\nu} + 4\sinh{r_1}^2$. We can then use the results above to obtain the metrological advantage, an example of which we have displayed on \figref{courbes}. In the first two cases, $I_F^{\text{opt}}>I_F^{\text{ref}}$, for all $\nu$. In the third case, $I_F^{\text{opt}} > I_F^{\text{ref}}$ if and only if $\nu>1$. If $\nu=1$, $I_F^{\text{opt}} = I_F^{\text{ref}}$: the FTQL can be attained, but not surpassed. This shows that Theorem 1 is also valid for most pure states, except one-mode displaced squeezed states with $V<0$ and $-4\frac{\lvert\gamma\rvert^2}{\nu}V \geq \frac{4\nu^2}{\nu^2+1}\big(\sinh^2(2r_1)-2\sinh^2(r_1)\big)$ (Which means that displacement and squeezing are roughly in the same direction, and displacement is large with respect to squeezing).

 \section{One-mode interferometry}
 \label{appendixE}

 We will now consider metrological advantage in one-mode phase estimation. A one-mode Gaussian state may be decomposed as:

\begin{equation}
\label{capacityonemode}
	\Dop(\bm{\gamma})\Rop(\phi)\Sop(r)\rop_{\text{th}}(\nu)(.)^{\dagger}
\end{equation}
where $\Rop$ represents one-mode phase-shifting and $\Sop$ one-mode squeezing. Here $\gamma=(\lvert\gamma\rvert e^{i\phi_{d}},\lvert\gamma\rvert e^{-i\phi_{d}})^T$. The QFI associated with phase estimation reads \cite{safranek_gaussian_2016}:

\begin{equation*}
	I_F^{\text{opt}}(\rop_{x})=4\frac{\nu^2}{\nu^2+1}\sinh^2(2r) + \frac{4\lvert\gamma\rvert^2}{\nu}\Big(e^{2r}\cos(\tilde{\phi})+e^{-2r}\sin(\tilde{\phi})\Big)
\end{equation*}
with $\tilde{\phi}=\phi_d+\phi_1$. Note that the only operations available are phase-shifts, which leave $\phi+\phi_d$ invariant: therefore we cannot optimize the precision any further. When $\nu=1$, this expression is equivalent to \eqref{specificase} with $a=b=\frac{\pi}{4}$. 

Inspired by the discussion in Appendix \ref{appendixD} above, let us consider the case $\tilde{\phi}=\frac{\pi}{2}$. If $$\frac{4\lvert\gamma\rvert^2}{\nu} > \frac{1}{1-e^{-2r}}\Big(4\frac{\nu^2}{\nu^2+1}\sinh(2r)^2-2\sinh(r)^2\Big)$$ (That is, the displacement is large with respect to squeezing) then $I_F^{\text{opt}}(\rop)-I_F^{\text{ref}} < 0$. In that case, the FTQL is not even attainable. 

\section{Interaction with ancilla}
\label{appendixF}

\begin{figure}
\captionsetup{width=.5\textwidth}
\centering
\includegraphics[scale=0.25]{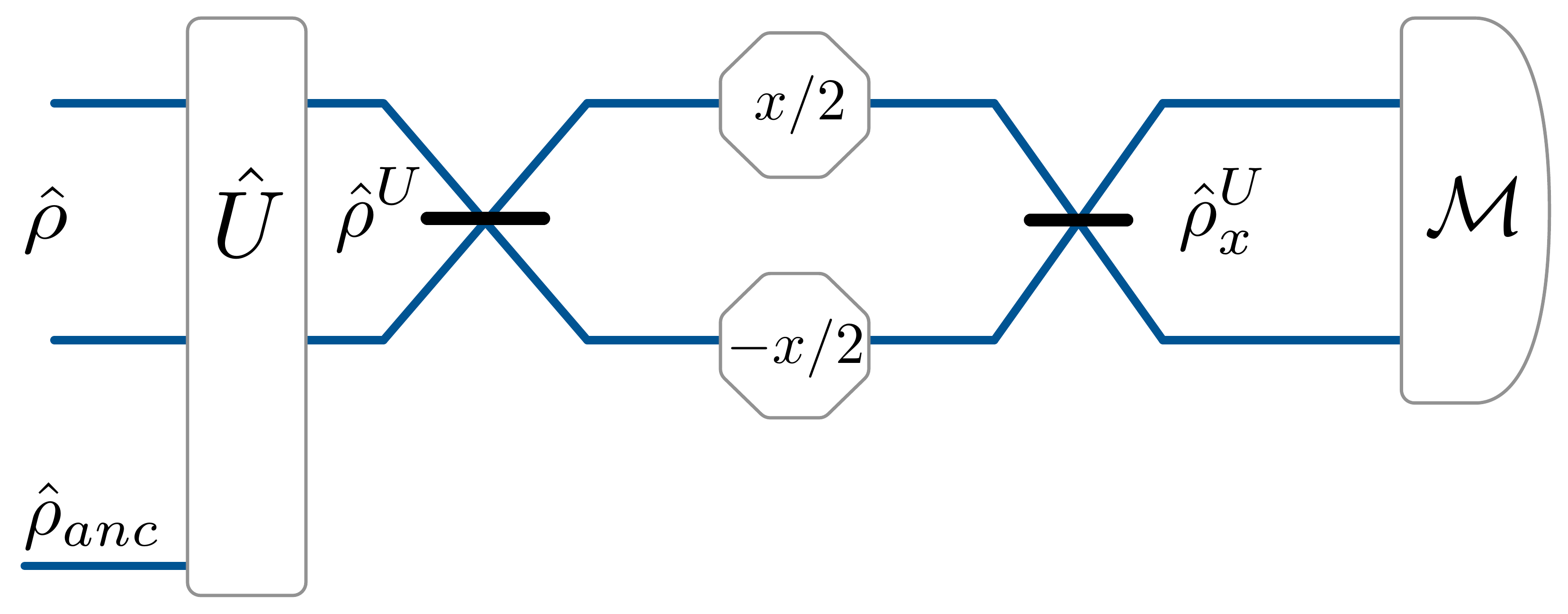}
\caption{Addition of ancilla. A displaced thermal state $\rop_{\text{anc}}$ is sent in a third mode; the three modes interacts via a controlled free unitary, then the ancilla is discarded. If the ancilla is a coherent state with a lot of photons, and $\Uop$ is just a mode-mixing operation between the ancilla and the first mode, this operation amounts to a displacement of the first mode.}\label{ancillainteract}
\end{figure}

We will consider the following operation: add a third mode occupied by a coherent state, put a beam-splitter between the first (say) and third modes, then trace out the third mode. This operation is schematically pictured in \figref{ancillainteract}.  
If the transmittivity of the beam-splitter is weak, and in the limit where the coherent state in the third mode is strong, this operation may be approximated as a displacement of the two-mode state. \\

Now, let us consider a state with $\theta=\psi=\alpha=r_2=0$, $\nu=1$, and $\phi_1=-\phi_{d1}+\frac{\pi}{2}$ (displaced squeezed state in mode 1 with displacement and squeezing orthogonal, thermal state in mode 2). Then we can use the results of Appendix \ref{appendixD} to write:
\begin{align}
\nonumber
	I_F(\rop^U) = & \sin^2(2b)\sin^2(2a)\Big[2\big(\sinh^2(2r_1)-2\sinh^2(r_1)\big) \\ \nonumber
	& + 4\lvert\gamma\rvert^2(e^{2r_1} - 1\big)\Big] + 4\lvert\gamma\rvert^2 + 4\sinh{r_1}^2
\end{align}
The precision is optimized by setting $a=b=\frac{\pi}{4}$. Then we have (using $I_F^{\text{ref}}=4\lvert\gamma\rvert^2 + 4 \sinh^2(r_1^2)$):

\begin{equation}
\label{one-modepureoptimized}
I_F^{\text{opt}}-I_F^{\text{ref}} = 2\big(\sinh^2(2r_1)-2\sinh^2(r_1)\big) + 4\lvert\gamma\rvert^2(e^{2r_1} - 1\big)
\end{equation}

This expression is obviously increasing with $\lvert\gamma\rvert$. Hence, if we take this state and let it interact with an ancilla as described above, we will be able to increase $\lvert\gamma\rvert$, and hence increase its metrological advantage.

\section{Renormalization of the metrological advantage}
\label{appendixG}

First of all, let us have a look at Eq.\eqref{eqdep}. The pure squeezing terms $Z$,$X$,$Y$ depend on $\nu$ as $\frac{\nu^2}{\nu^2+1}$; hence, they increase with $\nu$. The displacement-squeezing term $\frac{4\lvert\gamma\rvert^2}{\nu}$, by contrast, decreases with $\nu$.\\

Now, let us see whether we could obtain a monotonically decreasing quantity by renormalizing the advantage by the FTQL. More precisely, we may define $$\mathcal{\tilde B}_G=\frac{I_F(\rop_{x}^{\uvect})-\Big(4\frac{\langle \Nop\rangle+1}{\nu}-4\Big)}{\Big(4\frac{\langle \Nop\rangle+1}{\nu}-4\Big)}$$. (For mixtures of Gaussian states, we would do convex roof minimization over this quantity). Let us consider the case $\gamma=0$ for simplicity. We have then $\Big(4\frac{\langle \Nop\rangle+1}{\nu}-4\Big)=4(\sinh(r_1)^2+\sinh(r_2)^2)$. The FTQL is independent of $\nu$ in that case, which means the normalized and unnormalized advantages will have the same $\nu$-dependency. Thus, $\mathcal{\tilde B}_G$ can also increase with $\nu$. \\

We can also check how this renormalized quantity evolve with $\lvert\gamma\rvert$. Consider \eqref{specificase} when $\tilde{\phi}=\frac{\pi}{2}$ and $\nu=1$; the state is optimized by setting $a=b=\frac{\pi}{4}$. We obtain $I_F^{\text{opt}}=2\sinh^2(2r_1)+ 4\lvert\gamma\rvert^2e^{2r_1}$, and $$\frac{I_F^{\text{opt}}-I_F^{\text{ref}}}{I_F^{\text{ref}}} = \frac{2\sinh^2(2r_1)+ 4\lvert\gamma\rvert^2e^{2r_1}}{4\sinh^2(r_1)+ 4\lvert\gamma\rvert^2} - 1$$
It is not difficult to show that this function increases with $\lvert\gamma\rvert$ when $r_1$ is large. Hence, $\mathcal{\tilde B}_G$ can also be increased by the interaction with a free ancilla, which means it does not have all the properties of a resource monotone either.\\

\section{Separability of displaced thermal states}
\label{separability displtherm}

First, we consider two-mode displaced Fock states 

$\ket{\beta_1 p}\ket{\beta_2 q}=\Dop(\beta_1)\ket{p}\Dop(\beta_2)\ket{q}=\sum_{N}\ket{\beta_1 p}\ket{\beta_2 q}^{(N)}$

Here $\ket{\beta_1 p}\ket{\beta_2 q}^{(N)}$ is the component of $\ket{\beta_1 p}\ket{\beta_2 q}$ with a total average number of particle equal to $N$.

We will focus on the $N=2$ subspace. By direct computation, the state in this subspace reads:

\begin{widetext}
\begin{align}
	\ket{\beta_1 p}\ket{\beta_2 q}^{(2)}=\frac{\bar{\beta_1}^p\bar{\beta_2}^q}{\sqrt{p!q!}}e^{-\frac{\lvert\beta_1\rvert^2+\lvert\beta_2\rvert^2}{2}} \Big[ & \frac{(\beta_1)^2}{\sqrt{2}} \Big(1+\frac{2p}{\lvert\beta_1\rvert^2}+\frac{p(p-1)}{\lvert\beta_1\rvert^4}\Big) \ket{20} + \frac{(\beta_2)^2}{\sqrt{2}} \Big(1+\frac{2q}{\lvert\beta_2\rvert^2}+\frac{q(q-1)}{\lvert\beta_2\rvert^4}\Big)\ket{02}\\
	& + \beta_1\beta_2\Big(1+\frac{p}{\lvert\beta_1\rvert^2}\Big)\Big(1+\frac{q}{\lvert\beta_2\rvert^2}\Big)\ket{11} \Big]
\end{align}

\end{widetext}
where $\ket{n p}$ means n(p) photons in the first(second) mode. We have omitted a normalization factor to avoid cluttering. Next, we compute the expression for two-mode displaced thermal state $\rop_{th}(\beta_1,\beta_2)=\sum_{p,q}\frac{\Theta_1^p\Theta_2^q}{Z_1Z_2}\ket{\beta_1 p}\ket{\beta_2 q}\bra{\beta_1 p}\bra{\beta_2 q}$

where $\Theta_i=e^{-\frac{\hbar\omega_i}{kT_i}}$, and the $Z_i$ are partition functions. In the $N=2$ subspace, tedious but straightforward computations yield (up to normalization):
\begin{widetext}
\begin{align}
\label{expression_discoh}
	\rop_{th}(\beta_1,\beta_2)^{(2)}&=C\Big[\Lambda_1 \ket{20}\bra{20} + \Lambda_2 \ket{02}\bra{02} + \Omega \ket{11}\bra{11} + (\Upsilon_1 \ket{11}\bra{20} + \Upsilon_2 \ket{11}\bra{02} + \Xi\ket{20}\bra{02} + c.c)\Big]\\ \nonumber
	C&=\frac{1}{Z_1Z_2}e^{\lvert\beta_1\rvert^2(\Theta_1-1)+\lvert\beta_2\rvert^2(\Theta_2-1)}\\ \nonumber
	\Lambda_i&=\frac{1}{2}\Big(\lvert\beta_i\rvert^4(1+\Theta_i)^4+4\lvert\beta_i\rvert^2\Theta_i(1+\Theta_i)^2+2\Theta_i^2)\\ \nonumber
	\Upsilon_1&=\frac{\bar{\beta}_1\beta_2}{\sqrt{2}}(1+\Theta_1)(1+\Theta_2)\Big(\lvert\beta_1\rvert^2(1+\Theta_1)^2 + 2\Theta_1 \Big)\\ \nonumber
	\Xi&=\beta_1^2\bar{\beta}_2^2\frac{(1+\Theta_1)^2(1+\Theta_2)^2}{2}\\ \nonumber
	\Omega&=\Big(\lvert\beta_1\rvert^2(1+\Theta_1)^2+\Theta_1\Big)\Big(\lvert\beta_2\rvert^2(1+\Theta_2)^2+\Theta_2\Big)
\end{align}
\end{widetext}

 In the single particle basis ($\ket{\psi_1^a\psi_1^b}$,$\ket{\psi_1^a\psi_2^b}$, $\ket{\psi_2^a\psi_1^b}$, $\ket{\psi_2^a\psi_2^b}$), we can rewrite the density matrix as:

\begin{equation}
 	M=\begin{bmatrix}
 	\Lambda_1 & \frac{\Upsilon_1}{\sqrt{2}} & \frac{\Upsilon_1}{\sqrt{2}} & \Xi\\
 	\frac{\bar{\Upsilon}_1}{\sqrt{2}} & \frac{\Omega}{2} & \frac{\Omega}{2} & \frac{\Upsilon_2}{\sqrt{2}}\\ 
 	\frac{\bar{\Upsilon}_1}{\sqrt{2}} & \frac{\Omega}{2} & \frac{\Omega}{2} & \frac{\Upsilon_2}{\sqrt{2}}\\ 
 	\bar{\Xi} & \frac{\bar{\Upsilon}_2}{\sqrt{2}} & \frac{\bar{\Upsilon}_2}{\sqrt{2}} & \Lambda_2 
 	\end{bmatrix}
 \end{equation} 
To study the separability of this state, we used Peres-Horodecki criterion \cite{peres_separability_1996,horodecki_separability_1996}. We compute the partial transpose of the matrix, then rewrite it in the $\ket{\psi_1^a\psi_1^b},\frac{\ket{\psi_1^a\psi_2^b} + \ket{\psi_2^a\psi_1^b}}{2}, \ket{\psi_2^a\psi_2^b}, \frac{\ket{\psi_1^a\psi_2^b} - \ket{\psi_2^a\psi_1^b}}{2}$ basis. We obtain:
\begin{equation}
	\begin{bmatrix}
 	\Lambda_1 & \frac{\Upsilon_1+\bar{\Upsilon}_1}{2} & \frac{\Omega}{2} & \frac{\bar{\Upsilon}_1-\Upsilon_1}{2}\\ 
 	\frac{\Upsilon_1+\bar{\Upsilon}_1}{2} & \frac{\Omega}{2} + \frac{\Xi+\bar{\Xi}}{2} & \frac{\Upsilon_2+\bar{\Upsilon}_2}{2} & 0\\ 
 	\frac{\Omega}{2} & \frac{\Upsilon_2+\bar{\Upsilon}_2}{2} & \Lambda_2 & \frac{\bar{\Upsilon}_2-\Upsilon_2}{2}\\ 
 	\frac{\Upsilon_1-\bar{\Upsilon}_1}{2} & 0 & \frac{\Upsilon_2-\bar{\Upsilon}_2}{2} & \frac{\Omega}{2} - \frac{\Xi+\bar{\Xi}}{2}
 	\end{bmatrix}
\end{equation}
$\rop_{th}(\beta_1,\beta_2)^{(2)}$ is separable if and only if this matrix is positive.
In what follows, we will consider only the symmetric case $\Theta_1=\Theta_2=\Theta$ and $\beta_1=\beta_2=\beta$. This means $\Lambda_1=\Lambda_2=\Lambda$, $\Upsilon_1=\Upsilon_2=\Upsilon=\bar{\Upsilon}$, $\Xi=\bar{\Xi}$.

The matrix reduces to :
\begin{equation}
	\begin{bmatrix}
 	\Lambda & \Upsilon & \frac{\Omega}{2} & 0\\ 
 	\Upsilon & \frac{\Omega}{2} + \Xi & \Upsilon & 0\\ 
 	\frac{\Omega}{2} & \Upsilon & \Lambda & 0\\ 
 	0 & 0 & 0 & \frac{\Omega}{2} - \Xi
 	\end{bmatrix}
\end{equation}
We can immediately isolate the eigenvalue $\frac{\Omega}{2}-\Xi$. Using the expressions \eqref{expression_discoh}, this quantity is clearly positive. We note $y_1$, $y_2$, $y_3$ the other eigenvalues. Using the derivatives of the characteristic polynomial, we obtain: 
\begin{align}
	y_1+y_2+y_3 & =2\Lambda+\Xi+\frac{\Omega}{2}\\ \nonumber
	y_1y_2+y_3y_2+y_1y_3 & = \Lambda^2-\frac{\Omega^2}{4} + 2\Big(\Xi+\frac{\Omega}{2}\Big)\Lambda - 2\Lambda^2\\\nonumber
	y_1y_2y_3 & =\Lambda^2(\Omega-2\Lambda)+\Big(\Xi+\frac{\Omega}{2}\Big)\Big(\Lambda^2-\frac{\Omega^2}{4}\Big)\\\nonumber
	 & =\Big(\Lambda-\frac{\Omega}{2}\Big)\Big[\Big(\Xi+\frac{\Omega}{2}\Big)\Big(\Lambda+\frac{\Omega}{2}\Big)-2\Upsilon^2\Big]
\end{align}

Using \eqref{expression_discoh}, it is straightforward to show that these three expressions are positive. We can directly deduce that $y_1$, $y_2$ and $y_3$ are postive, which means the $N=2$ substate of a symmetric displaced thermal state is particle-separable.

\end{document}